\documentclass[aps,prd,showpacs,showkeys,amsmath,amssymb]{revtex4}
\usepackage{graphicx}

\begin{document}

\title{Pentaquarks}
\author{Shi-Lin Zhu}
\email{zhusl@th.phy.pku.edu.cn}
\affiliation{%
Department of Physics, Peking University, BEIJING 100871, CHINA}

\date{\today}

\begin{abstract}

Since LEPS collaboration reported the first evidence of $\Theta^+$
pentaquark in early 2003, eleven other experimental groups have
confirmed this exotic state while many other groups didn't see any
signal. If this state is further established by future high
statistical experiments, its discovery shall be one of the most
important events in hadron physics for the past three decades.
This exotic baryon with such a low mass and so narrow a width
imposes a big challenge to hadron theorists. Up to now, there have
appeared more than two hundred theoretical papers trying to
interpret this charming state. I will review some important
theoretical developments on pentaquarks based on my biased
personal views.

\end{abstract}


\maketitle

\pagenumbering{arabic}

\newpage

\begin{center}
{\bf\Large {Table of contents}}
\end{center}
\vspace{0.5cm}

\begin{itemize}

\item I. Quark Model and Exotic Hadron Search

\item II. Pentaquarks: Discovery or Fluctuations?

\item III. Group Theory Of Pentaquarks

\item IV. Theoretical Models of Pentaquarks

              \begin{itemize}

              \item  Chiral soliton model

              \item  Jaffe and Wilczek's Diquark model

              \item  Other clustered quark model

              \item  Non-clustered quark model

              \item  QCD sum rules

              \item  Heptaquarks

              \end{itemize}

\item V. Narrow width puzzle

\item VI. Pentaquark flavor wave functions, masses and magnetic
          moments

\item VII. Heavy flavored pentaquarks

\item VIII. Chiral Lagrangian formalism for pentaquarks

           \begin{itemize}

              \item Mass splitting

              \item Selection rules from octet pentaquark decays

              \end{itemize}

\item IX. Diquarks, pentaquarks and dibaryons

\item X. Summary

\end{itemize}

\newpage

\section{Quark Model and Exotic Hadron Search}\label{sec1}

Quantum Chromodynamics (QCD) is believed to be the underlying
theory of the strong interaction, which has three fundamental
properties: asymptotic freedom, color confinement, approximate
chiral symmetry and its spontaneous breaking. In the high energy
regime, QCD has been tested up to $1\%$ level. In the low energy
sector, QCD is highly nonperturbative due to the non-abelian
SU$_c$(3) color group structure. It is very difficult to calculate
the whole hadron spectrum from first principles in QCD. With the
rapid development of new ideas and computing power, lattice gauge
theory may provide the final solution to the spectrum problem in
the future. But now, people have just been able to understand the
first orbital and radial excitations of the nucleon with lattice
QCD in the baryon sector \cite{liu}.

Under such a circumstance, various models which are QCD-based or
incorporate some important properties of QCD were proposed to
explain the hadron spectrum and other low-energy properties. Among
them, it is fair to say that quark model has been the most
successful one. It is widely used to classify hadrons and
calculate their masses, static properties and low-energy reactions
\cite{isgur}. According to quark model, mesons are composed of a
pair of quark and anti-quark while baryons are composed of three
quarks. Both mesons and baryons are color singlets. Most of the
experimentally observed hadrons can be easily accommodated in the
quark model. Any state with the quark content other than $q\bar q$
or $q q q$ is beyond quark model, which is termed as
non-conventional or exotic. For example, it is hard for
$f_0(980)/a_0(980)$ to find a suitable position in quark model.
Instead it could be a kaon molecule or four quark state
\cite{pdg}.

However, besides conventional mesons and baryons, QCD itself does
not exclude the existence of the non-conventional states such as
glueballs ($gg, ggg, \cdots$), hybrid mesons ($q\bar q g$), and
other multi-quark states ($qq\bar q \bar q$, $qqqq\bar q$,
$qqq\bar q \bar q \bar q$, $qqqqqq, \cdots$). In fact, hybrid
mesons can mix freely with conventional mesons in the large $N_c$
limit \cite{cohen}. In the early days of QCD, Jaffe proposed the H
particle \cite{jaffeold} with MIT bag model, which was a six quark
state. Unfortunately it was not found experimentally.

In the past years there have accumulated some experimental
evidence of possible existence of glueballs  and hybrid mesons
with exotic quantum numbers like $J^{PC}=1^{-+}$ \cite{pdg}.
Recently BES collaboration observed a possible signal of a proton
anti-proton baryonium in the $J/\Psi$ radiative decays \cite{bes}.
But none of these states has been pinned down without controversy
until the surprising discovery of pentaquarks by LEPS
collaboration \cite{leps}.

\section{Pentaquarks: Discovery or Fluctuations?}\label{sec2}

\begin{figure}
\scalebox{0.5}{\includegraphics{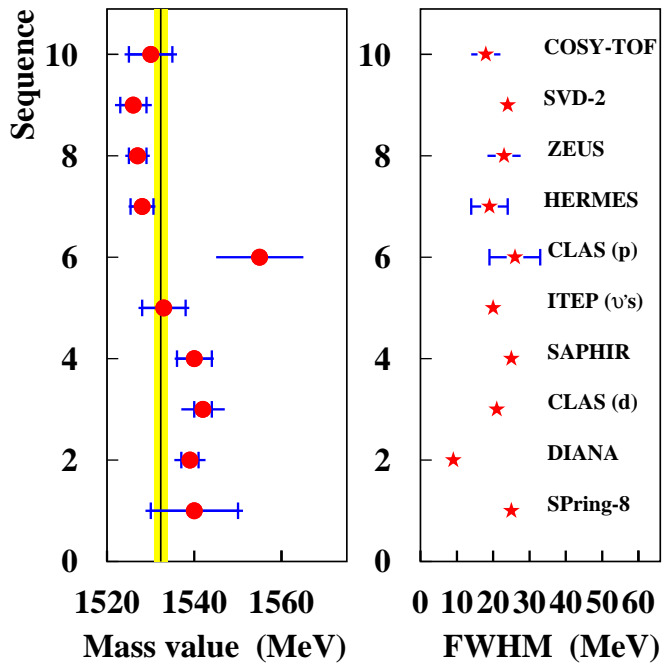}}
\vspace{0.5cm}\caption{The mass and decay width of $\Theta^+$
pentaquark from various experimental measurements. } \label{mao}
\end{figure}

Early last year LEPS Collaboration at the SPring-8 facility in
Japan observed a sharp resonance $\Theta^+$ at $1.54\pm 0.01$ GeV
with a width smaller than 25 MeV and a statistical significance of
$4.6\sigma$ in the reaction $\gamma n \to K^+ K^- n$ \cite{leps}.
This resonance decays into $K^+ n$, hence carries strangeness
$S=+1$. Later, many other groups have claimed the observation of
this state \cite{diana,clas,saphir,itep,clasnew,hermes,svd,cosy,
Yerevan,zeus,forzeus}. All known baryons with $B=+1$ carry
negative or zero strangeness. Such a resonance is clearly beyond
the conventional quark model with the minimum quark content
$uudd\bar s$. Now it's called $\Theta^+$ pentaquark in literature.
A compilation of $\Theta^+$ mass and decay width is presented in
Figure (\ref{mao}).

NA49 Collaboration announced evidence for the existence of a new
narrow $\Xi^- \pi^-$ baryon resonance $\Xi^{--}_5$ with mass of
$(1.862\pm 0.002) $ GeV and width below the detector resolution of
about 0.018 GeV in proton-proton collisions at $\sqrt{s}=17.2$ GeV
\cite{na49}. The quantum number of this state is $Q=-2, S = -2, I
= 3/2$ and its quark content is $(d s d s \bar u)$. They also
observed signals for the $Q=0$ member of the same isospin quartet
with a quark content of $(d s u s \bar d)$ in the $\Xi^- \pi^+$
spectrum. The corresponding anti-baryon spectra also show
enhancements at the same invariant mass. H1 Collaboration claimed
the discovery of a heavy pentaquark around 3099 MeV with the quark
content $udud\bar{c}$ \cite{H1}. Very recently, STAR collaboration
at RHIC observed a narrow peak at $1734\pm 0.5\pm 5$ MeV in the
$\Lambda K_s^0$ invariant mass which was interpreted as an
$I={1\over 2}$ pentaquark \cite{star-rhic}. However its
antiparticle was not observed yet.

There is preliminary evidence that the $\Theta^+$ is an iso-scalar
because no enhancement was observed in the $pK^+$ invariant mass
distribution \cite{saphir,clasnew,hermes,forzeus}. The third
component of its isospin is $I_z=0$. So the $\Theta^+$ pentaquark
is very probably an iso-scalar if it is a member of the
anti-decuplet. At present, the possibility of this state being a
member of another multiplet is not completely excluded. Hence its
total isospin is probably zero. Most of the theoretical models
assume that $\Theta^+$ is in $SU(3)_f$ ${\bf\bar{10}}$
representation. All the other quantum numbers including its
angular momentum and parity remain undetermined. However, most of
theoretical work postulated its angular momentum to be one half
because of its low mass. But the possibility of $J={3\over 2}$
still can not be excluded completely.

It is important to point out that many other experimental groups
reported negative results \cite{bes1,hera-b,rhic}. For example,
the existence of $\Xi^{--}_5$ is still under debate \cite{doubt}.
Compared with 1640 $\Xi^-$ candidates produced in proton proton
collisions in NA49's analysis, WA89 collaboration found no signal
of $\Xi^{--}$ pentaquark with $676000$ $\Xi^-$ candidates in their
data sample \cite{wa89}. A long list of experiments yielding
negative results including those unpublished can be found in Ref.
\cite{long}. Although the $\Theta^+$ pentaquark has been listed as
a three-star resonance in the 2004 PDG, its existence is still not
completely established.

\section{Group Theory Of Pentaquarks}\label{sec3}

One can use some textbook group theory to write down the
pentaquark wave functions in the framework of quark model. Because
of its low mass, high orbital excitation with $L\ge 2$ seems
unlikely. Pauli principle requires the totally anti-symmetric wave
functions for the four light quarks. Since the anti-quark is in
the $[11]_C$ representation, the four quark color wave function is
$[211]_C$.

With $L=0$, hence $P=-$, the 4q spatial wave function is
symmetric, i.e., $[4]_O$. Their $SU(6)_{FS}$ spin-flavor wave
function must be $[31]_{FS}^{210}$ which contains $[22]_F^6 \times
[31]_S^3$ after decomposition into $SU(3)_F\times SU(2)_S$
\cite{strot,g1,g2,buccella}. Here 210, 6 etc is the dimension of
the representation. When combined with anti-quark, we get
$\left([33]_F^{10}+[21]_F^8\right)\times
\left([41]_S+[32]_S\right)$, which is nothing but $\left({\bar
{10}}_F+8_F\right)\times \left(({3\over 2})_S+({1\over
2})_S\right)$ in terms of more common notation. The total angular
momentum of the four quarks is one. The resulting exotic
anti-decuplet is always accompanied by a nearly degenerate octet.
Their angular momentum and parity is either $J^P={3\over 2}^-$ or
${1\over 2}^-$.

With $L=1$ and $P=+$, the four quark $SU(6)_{OFS}$
space-spin-flavor wave function must be $[31]_{OFS}$. Only this
representation can combine with $[211]_C$ color wave function to
ensure that the 4q total wave functions are anti-symmetric. The 4q
orbital wave function is $[31]_O$. There are four $SU(6)_{FS}$
wave functions $[4]_{FS}, [31]_{FS}, [22]_{FS}, [211]_{FS}$ which
allow the 4q total wave function to be anti-symmetric and lead to
the octet and exotic anti-decuplet \cite{g2,buccella}.

\section{Theoretical Models of Pentaquarks}\label{sec4}

We are still unable to solve the hadron spectrum starting from
first principles of QCD analytically. Lattice QCD simulation may
play a very important role eventually. But right now, lattice
simulation of pentaquarks by several groups has not converged yet.
For example, one lattice calculation favors positive parity for
pentaquarks \cite{chiu} while two previous lattice simulations
favor negative parity \cite{lattice}. The most recent lattice
simulation did not observe any bound pentaquark state in either
$I=0, J^P={1\over 2}^\pm$ or $I=1, J^P={1\over 2}^\pm$
\cite{liukf}. We will not discuss lattice results in this review.
Interested readers are referred to the above references.

In the field of pentaquarks, there are three outstanding pending
issues: the $\Theta^+$ parity, the dynamical generation of its low
mass and its narrow width. Up to now theoretical papers in the
literature can roughly be classified into two types according to
their assumption of $\Theta^+$ parity. Theoretical approaches
advocating positive parity include the chiral soliton model (CSM)
\cite{diak}, the diquark model \cite{jaffe}, some quark models
\cite{lipkin,riska,hosaka,carl,liuyx}, a lattice calculation
\cite{chiu}. On the other hand, some other theoretical approaches
tend to favor negative parity such as two lattice QCD simulations
\cite{lattice}, QCD sum rule approaches \cite{zhu,qsr}, several
quark model study \cite{zhang,carlson,wu}, and proposing stable
diamond structure for $\Theta^+$ \cite{song}. There are many
proposals to measure $\Theta^+$ parity either through doubly
polarized proton proton scattering \cite{thomas} or
photo-production processes \cite{proposal,prod}. Now most of
photo-production papers assume that pentaquarks are produced from
the following four processes: s-channel, u-channel, t-channel
through charged meson exchange and the contact seagull diagram. If
such a mechanism is correct, then the production cross section of
producing a positive parity $\Theta^+$ pentaquark is one or two
orders larger than that of producing a negative parity $\Theta^+$.
Moreover, the inclusion of charge vector meson in the t-channel
increases the cross section significantly \cite{prod}.

Since LEPS's first experimental paper, there have appeared more
than two hundred theoretical papers on pentaquarks. It's both
impossible and unnecessary to exhaust them in the present short
review. In the following, I focus on several models to illustrate
these pending issues.

\subsection{Chiral soliton model}

Theoretical study of pentaquark states dated back to the early
days of QCD using MIT bag model \cite{strot}. The recent interest
in the $\Theta^+$ pentaquark started with the prediction of its
mass, width and reaction channel from the chiral soliton model
(CSM) in Ref. \cite{diak} a few years ago. Diakonov et al.
proposed the possible existence of the $S=1$ $J^P={1\over 2}^+$
resonance at $1530$ MeV with a width less than 15 MeV using the
chiral soliton model \cite{diak}, which partly motivated the
recent experimental search of this particle. The authors argued
that $\Theta^+$ is the lightest member of the anti-decuplet
multiplet in the third rotational state of the chiral soliton
model. Assuming that the $N(1710)$ is a member of the
anti-decuplet, $\Theta^+$ mass is fixed with the symmetry
consideration of the model.

However this prediction is very sensitive to the model inputs
\cite{diak,mp,ellis}. For example, either adopting the commonly
used value $45$ MeV for the $\sigma$-term or identifying $N(1710)$
as a member of the anti-decuplet will lead to a $\Xi_5^{--}$
pentaquark mass $210$ MeV higher than that observed by NA49
Collaboration \cite{diak}. An updated reanalysis indicates
$\Theta^+$ mass is really quite low within the framework of the
chiral soliton model: 1430 MeV $\le m_{\Theta^+} \le$ 1660 MeV
\cite{ellis}. For example, with the new larger value $(79\pm 7)$
MeV and $(64\pm 7)$ MeV from two recent analysis \cite{sigmaterm}
for the $\sigma$-term, a fairly good description of both
$\Theta^+$ and $\Xi^{--}$ masses is possible \cite{ellis}. In CSM,
the narrow width comes from the cancellation between the coupling
constants in the leading order, next-leading order and
next-next-leading order large $N_c$ expansion.

However, the theoretical foundation of the treatment of the
pentaquarks in the chiral soliton model is challenged by the large
$N_c$ formalism in Refs. \cite{cohen2}. From the large $N_c$
consistency consideration, Cohen found that predictions for a
light collective $\Theta^+$ baryon state (with strangeness +1)
based on the collective quantization of chiral soliton models are
shown to be inconsistent with large $N_c$ QCD since collective
quantization is legitimate only for excitations which vanish as
$N_c \to \infty$.

In the large $N_c$ limit, the mass splitting between the $\Delta$
decuplet and the nucleon octet is ${\cal O}(1/N_c)$ while the
excitation energy of the vibration mode of the hedgehog is ${\cal
O}(1)$. Hence the rotational degree of freedom decouples from the
vibration mode in the large $N_c$ limit. The collective
quantization is valid for the octet and decuplet. The predictions
for these two multiplets from the chiral soliton model is
reliable.

In contrast, the mass splitting between the anti-decuplet and
octet is also ${\cal O}(1)$. In other words, the rotation and
vibration motions are not orthogonal. They will mix with each
other, which invalidates the collective quantization of the
rotational degrees of freedom. Therefore, Cohen concluded that the
prediction of $\Theta^+$ properties based on collective
quantization of CSM was fortuitous \cite{cohen2}. Large $N_c$
formalism doesn't predict the existence of $\Theta^+$ pentaquark.
However, large $N_c$ formalism may still be a useful tool to
relate properties of different pentaquarks such as their mass
splitting and decay width once one {\sl assumes} the existence of
pentaquarks \cite{cohen2,jenkins}. If $\Theta^+$ is established by
experiments, then one may use large $N_c$ expansion technique to
predict the existence of the other members and their properties
within the same multiplet.

The relationship between the bound state and the SU(3) rigid
rotator approaches to strangeness in the Skyrme model was
investigated in \cite{princeton}. It was found that the exotic
state may be an artifact of the rigid rotator approach to the
Skyrme model for large $N_c$ and small $m_K$.

\subsection{Jaffe and Wilczek's Diquark model}

Since early last year, there appeared many theoretical papers
trying to interpret these exotic states. Among them, Jaffe and
Wilczek's (JW) diquark model is a typical one \cite{jaffe}. In
their model, the $\Theta^+$ pentaquark is composed of a pair of
diquarks and a strange anti-quark. The flavor anti-decuplet is
always accompanied by an octet which is nearly degenerate and mix
with the decuplet.

Jaffe and Wilczek's (JW) proposed that there exists strong
correlation between the light quark pair when they are in the
anti-symmetric color $({\bar 3}_c)$, flavor $({\bar 3}_f)$,
isospin $(I=0)$ and spin $(J=0)$ configuration \cite{jaffe}. The
lighter the quarks, the stronger the correlation, which helps the
light quark pair form a diquark. For example, the ud diquark
behaves like a scalar with positive parity. Such correlation may
arise from the color spin force from the one gluon exchange or the
flavor spin force induced by the instanton interaction.

They proposed that pentaquark states are composed of two scalar
diquarks and one anti-quark. Diquarks obey Bose statistics. Each
diquark is in the antisymmetric color $\bf{\bar{3}}$ state. The
spin wave function of the two quarks within each scalar diquark is
antisymmetric while the spatial part is symmetric. Pauli principle
requires the total wave function of the two quarks in the diquark
be anti-symmetric. Thus the flavor wave function of the two quarks
in the diquark must be antisymmetric, i.e, the diquark is in the
flavor $\bf{\bar3_F}$ state.

The color wave function of the two diquarks within the pentaquark
must be antisymmetric $\bf{3}_C$ so that the resulting pentaquark
is a color singlet. In order to accommodate the exotic $\Theta^+$
pentaquark, Jaffe and Wilczek required the flavor wave function of
the diquark pair to be symmetric ${\bf {\bar 6_f}}$: $[ud]^2$,
$[ud][ds]_+$, $[su]^2$, $[su][ds]_+$, $[ds]^2$, and $[ds][ud]_+$.
Simple group theory tells us that ${\bf {\bar 6_f}}\times {\bf
{\bar 3_f}}={\bf {{\bar 10}_f}}+ {\bf 8_f}$. In other words, the
pentaquark anti-decuplet is always accompanied by an octet which
is nearly degenerate and probably mixes with the anti-decuplet.
Bose statistics demands symmetric total wave function of the
diquark-diquark system, which leads to the antisymmetric spatial
wave function with one orbital excitation. The resulting
anti-decuplet and octet pentaquarks have $J^P={1\over 2}^+,
{3\over 2}^+$.

In general, the lower the angular mentum, the lower the mass. So
the $J={3\over 2}$ pentaquark will be heavier than its $J={1\over
2}$ partner. But their mass difference is not expected to larger
than $300$ MeV if we could rely on the past experience with the
$\Delta$ and nucleon mass splitting. If the $\Theta^+$ pentaquark
does exist and the diquark model is correct, then its $J={3\over
2}$ pentaquark partner should also be reachable by future
experiments.

The members of pentaquark anti-decuplet are $P_{333}=\Theta^+$, $
P_{133}=\frac{1}{\sqrt{3}}N^0_{10}$,
$P_{233}=-\frac{1}{\sqrt{3}}N^+_{10}$, $P_{113}
=\frac{1}{\sqrt{3}}\Sigma^-_{10}$,
$P_{123}=-\frac{1}{\sqrt{6}}\Sigma^0_{10}$,
$P_{223}=\frac{1}{\sqrt{3}}\Sigma^+_{10}$,
$P_{111}=\Xi^{--}_{10}$, $P_{112} =-\frac{1}{\sqrt{3}}\Xi^-_{10}$,
$P_{122}=\frac{1}{\sqrt{3}}\Xi^0_{10}$ and $P_{222}=-\Xi^+_{10}$.

Jaffe and Wilczek pointed out that one of the decay modes
$\Xi_5\to \Xi^\ast + \pi$ observed by NA49 \cite{na49} signals the
existence of an additional octet around $1862$ MeV together the
anti-decuplet since the latter can not decay into a decuplet and
an octet in the $SU(3)_f$ symmetry limit. If further confirmed,
this experiment poses a serious challenge to the chiral soliton
model because there is no baryon pentaquark octet in the
rotational band in this model. But it may be hard to exclude it
since there always exist excited vibrational octet modes. These
modes can not be calculated rigourously within the chiral soliton
model \cite{ellis}.

However, in JW's model $\Theta^+$ is not the lightest pentaquark
as in the chiral soliton model. The ideal mixing between the octet
and anti-decuplet will split the spectrum and produce two
nucleon-like states. Using the $\Theta^+$ mass as input, the
higher one $N_s$ is around 1710 MeV with a quark content $qqq \bar
s s$ where $q$ is the up or down quark. The lighter one $N_l$ is
lower than the $\Theta^+$ pentaquark with a quark content $qqq
\bar q q$. Jaffe and Wilczek identified $N_l$ as the well-known
Roper resonance $N(1440)$ \cite{jaffe}, which is a very broad
four-star resonance with a width around $(250 \sim 450)$ MeV.
However, it will be very demanding to explain Roper's large decay
width and $\Theta^+$'s extremely narrow width simultaneously in a
natural way. The problem with the identification of the ideally
mixed positive parity pentaquarks with the $N(1710)$ and $N(1440)$
is discussed extensively in \cite{wrong}.

If Jaffe and Wilczek's diquark picture for $\Theta_5^+$ pentaquark
is correct, we pointed out \cite{zhanga} that lighter pentaquarks
can be formed if the two scalar diquarks are in the antisymmetric
$SU(3)_F$ $\bf 3$ representation: $[ud][su]_-$, $[ud][ds]_-$, and
$[su][ds]_-$, where
$[q_1q_2][q_3q_4]_-=\sqrt{\frac{1}{2}}([q_1q_2][q_3q_4]-[q_3q_4][q_1q_2])$.
No orbital excitation is needed to ensure the symmetric total wave
function of two diquarks since the spin-flavor-color part is
symmetric. The total angular momentum of these pentaquarks is
$\frac{1}{2}$ and the parity is negative. There is no accompanying
$J={3\over 2}$ multiplet. The two diquarks combine with the
antiquark to form a $SU(3)_F$ pentaquark octet and singlet: ${\bar
3}_F \otimes 3_F = \bf{8}_F$ $\oplus$ $\bf{1}_F$. Similar
mechanism has been proposed to study heavy pentaquarks with
negative parity and lighter mass than $\Theta_{c,b}$ in
\cite{wise}.

Because there is no orbital excitation within these nine
pentaquarks, their masses are lower than the anti-decuplet and the
accompanying octet with positive parity. Their masses range
between 1360 MeV and 1540 MeV according to our calculation using
the same mass formula in \cite{jaffe}. These states are close to
the L=1 orbital excitations of the nucleon octet. The mixing
between the pentaquark states and orbital excitations is expected
to be small since their spatial wave functions are very different.

According to our calculation, two of the $J^P={1\over 2}^-$ octet
pentaquark members $p_8, n_8$ lie 22 MeV below the $p\eta_0$
threshold and 228 MeV below the $\Sigma K$ threshold. The
"fall-apart" decay mechanism forbids $p_8$ to decay into one
nucleon and one pion. Although their interaction is of S-wave,
lack of phase space forbids the strong decays $p_8\to p\eta_0,
\Sigma K$ to happen.

For $p_8$, the only kinematically strong decays are S-wave $p_8\to
p\pi$, P-wave $p_8 \to N \pi \pi$ and S-wave $p_8\to N \pi\pi\pi$
decays where $N$ is either a proton or neutron. All these decay
modes involve the annihilation of a strange quark pair and violate
the the "fall-apart" mechanism. Hence the width is expected to be
small. Both $p_8$ and $n_8$ may be narrow resonances.

It is interesting to note there are three negative-parity
$\Lambda$ particles within the range between 1400 MeV and 1540 MeV
if Jaffe and Wilczek's diquark model is correct. One of them is
the well established $\Lambda (1405)$. $\Lambda (1405)$ is only 30
MeV below kaon and nucleon threshold. Some people postulated it to
be a kaon nucleon molecule \cite{pdg}. We propose that there is
another intriguing possibility of interpreting $\Lambda (1405)$ as
the candidate of $J^P={1\over 2}^-$ pentaquark. The other
$J^P={1\over 2}^-$ pentaquark and the corresponding L=1 singlet
$\Lambda$ particle may have escaped detection so far.

The discovery of nine additional negative-parity baryons in this
mass range will be strong evidence supporting the diquark model.
Then one has to consider the possibility of treating diquarks as
an effective degree of freedom as constituent quarks when dealing
with hadron spectrum. One possible consequence is the appearance
of many exotic mesons and over-population of meson states with
non-exotic quantum numbers below 2 GeV. On the other hand, if
future experimental searches fail to find any evidence of these
additional states with negative parity, one has to re-evaluate the
relevance of the diquark picture for the pentaquarks.

\subsection{Other clustered quark model}

An even simpler picture came from Karliner and Lipkin's diquark
triquark model \cite{lipkin}. In their model the two color
non-singlet clusters are kept apart by the P-wave angular momentum
barrier. Hence the color magnetic interaction occurs within two
clusters only. Between two clusters the color electric forces bind
them into a color singlet. The angular momentum barrier prevents
them from rearranging into the usual $KN$ system. The presence of
P-wave ensures the overall $\Theta^+$ parity is positive.

The two quarks in the triquark are in the symmetric ${\bf 6_c}$
representation. They couple with the antiquark to form an
$SU(3)_c$ triplet ${\bf 3}_c$.  The two quarks are in the
anti-symmetric flavor ${\bf \bar{3}_f}$ representation and the
triquark is in the symmetric ${\bf \bar 6}_f$. Thus the spin wave
function of the two quarks is symmetric. The spin of the triquark
is one half.

The direct product of the ${\bf \bar 3_f}$ of diquark and the
${\bf \bar 6_f}$ of triquark leads to ${\bf\bar{10}_f}$ and ${\bf
8_f}$ pentaquarks. There is one orbital angular momentum $L=1$
between the diquark and the triquark. The resulting $J^P$ of the
pentaquark is ${1\over 2}^+$ or ${3\over 2}^+$.

Based on the color-magnetic interaction, a recent paper analyzed
the triquark cluster including the explicit SU(3) symmetry
breaking \cite{hogaasen}. With the reasonable parameters $C_{ij}$
from the analysis of the conventional baryons, the $\Theta^+$ mass
is found to be around 1520 MeV even without the orbital excitation
energy. It's very demanding to reproduce the low mass of the
$\Theta^+$ pentaquark!

Shuryak and Zahed's (SZ) suggested that the pentaquark mass be
lower by replacing one scalar diquark with one tensor diaquark in
JW's model \cite{shuryak}. But their $\Theta^+$ pentaquark is in
the $27$ representation and $\Theta^+$ is inevitably accompanied
by a $\Theta^{++}$ in their model.

\subsection{Non-clustered quark model}

Stancu and Riska studied the stability of the strange pentaquark
state assuming a flavor-spin hyperfine interaction between quarks
in the constituent quark model \cite{riska}. They suggested that
the lowest lying p-shell pentaquark state with positive parity
could be stable against strong decays if the spin-spin interaction
between the strange antiquark and up/down quark was strong enough
\cite{riska}.

In Ref. \cite{hosaka} Hosaka emphasized the important role of the
hedgehog pion in the strong interaction dynamics which drives the
formation of the $\Theta^+$ particle and the detailed energy level
ordering.

In the adiabatic approximation, Zhang et al employed the chiral
quark model to make a systematic comparison study of the
$\Theta^+$ pentaquark mass with both parities \cite{zhang}. Their
model Lagrangian includes the one pion exchange force, one gluon
exchange force and the linear confinement. First they reproduced
known hadron spectroscopy and nucleon-nucleon, nucleon-hyperon
scattering phase shift. In this way most of their model parameters
were fixed. Then they extended their discussion to the pentaquark
system. They found that the pentaquark in the I=0 channel in the
anti-decuplet representation is always lowest for both
$J^P={1\over 2}^+$ and $J^P={1\over 2}^-$. Without interaction
between four light quarks and strange anti-quark, the pentaquark
with $J^P={1\over 2}^+$ lies lower than that of $J^P={1\over 2}^-$
as pointed out by Stancu and Riska \cite{riska}. However, the
$J^P={1\over 2}^+$ pentaquark mass is about 400 MeV above
$\Theta^+$. With interaction (especially annihilation effects)
between 4q and $\bar s$, the $J^P={1\over 2}^-$ pentaquark is
lower, with a mass (150-300) MeV above $\Theta^+$. It's fair to
say that it is still very demanding to dynamically generate the
$\Theta^+$ pentaquark with a low mass of $1530$ MeV within the
framework of the non-clustered quark models.

Wang et al. tried three models to calculate $\Theta^+$ pentaquark
mass \cite{wang}. First they treated $\Theta^+$ as a pure
$uudd\bar s$ five quark state including channel coupling. Their
numerical results indicate the hidden color channel effects and
coupled channel effects play a vital role in reducing $\Theta^+$
mass. However the S-wave state is always lower than P-wave one.
The second model they used is the quark delocalization, color
screening model. The third one mimics the diquark color
configuration. They found a pure repulsive effective action for
the $I=1$ $KN$ channel, hence excluding $I=1$ possibility for
$\Theta^+$. In the adiabatic approximation, the $\Theta^+$ mass of
$1615$ MeV is obtained in the $I=0$ S-wave $KN$ channel. In their
model the P-wave attraction is not sufficient to overcome the
kinetic increase hence the positive parity pentaquark state lies
higher than the negative parity one. It is emphasized that the
possibility of $\Theta^+$ being a $I=0J^P={1\over 2}^-$ state has
not been ruled out yet \cite{wang}.

From previous quark model experience, one would naively expect the
strange pentaquark mass with positive parity to be around $4M_u +
M_s + 400\sim 2050$ MeV, where $M_u=300$ MeV and $M_s=450$ MeV are
roughly the up and strange quark constituent mass. The additional
$\sim 400$ MeV comes from one orbital excitation for the
positive-parity $\Theta^+$. If future experiments really establish
the $\Theta^+$ parity to be positive, then we have to conclude
that there must exist some kind of unknown and strong correlation
between the light quarks in the pentaquark system which helps
lower its mass significantly.

\subsection{QCD sum rules}

The method of QCD sum rules (QSR) can not solve color confinement.
Instead, QSR incorporates two basic properties of QCD as its
starting point: confinement and approximate chiral symmetry and
its spontaneous breaking. One considers a correlation function of
some specific interpolating currents with the proper quantum
numbers as in lattice QCD simulation and calculates the correlator
perturbatively starting from high energy region. Then the
resonance region is approached where non-perturbative corrections
in terms of various condensates gradually become important. Using
the operator product expansion, the spectral density of the
correlator at the quark gluon level can be obtained in QCD. On the
other hand, the spectral density can be expressed in term of
physical observables like masses, decay constants, coupling
constants etc at the hadron level. With the assumption of quark
hadron duality these two spectral densities can be related to each
other. In this way one can extract hadron masses etc. For the past
decades QCD sum rules has proven to be a very powerful and
successful non-perturbative method \cite{svz,reinders}.

The $\Theta^+$ pentaquark mass was first studied using the
interpolating current \cite{zhu}
\begin{eqnarray}\label{cu3} \nonumber
\eta_0(x)={1\over \sqrt{2}} \epsilon^{abc} [u^T_a(x) C\gamma_5
d_b (x)] \{ u_e (x)\\
\times {\bar s}_e (x) i\gamma_5 d_c(x) - \left( u\leftrightarrow
d\right) \}
\end{eqnarray}
where $\left( C\gamma_5\right)^T =-C\gamma_5$ ensures the isospin
of the up and down quark pair inside the first bracket to be zero.
The anti-symmetrization in the second bracket ensures that the
isospin of the other up and down quark pair is also zero.

The following correlation function was considered in Ref.
\cite{zhu}
\begin{equation}\label{cor-1}
i\int d^4 x e^{ipx} \langle 0|T\left (\eta_j (x), {\bar \eta}_j
(0) \right )|0\rangle\ = \Pi (p) {\hat p} + \Pi' (p )
\end{equation}
where $\bar \eta =\eta^\dag \gamma_0$ and ${\hat p} =p_\mu \cdot
\gamma^\mu$. The spectral density of $\Pi (p)$ reads
\begin{eqnarray}\label{spectral}\nonumber
\rho_0={3 s^5\over  4^8 7! \pi^8}+{s^2\over 1536 \pi^4} [ {5\over
12} \langle \bar q q\rangle^2  +{11\over 24}\langle \bar
qq\rangle\langle \bar ss\rangle ]\\ \nonumber +[{7\over
432}\langle \bar q q\rangle^3 \langle \bar s s\rangle +{1\over
864}\langle \bar q q\rangle^4] \delta (s)
\end{eqnarray}
after invoking the factorization approximation for the multi-quark
condensates.

The mass of $\Theta^+$ pentaquark is
\begin{eqnarray}
M_0^2 =\frac{\int_{m_s^2}^{s_0} d s e^{-{s/T^2}} \rho'
(s)}{\int_{m_s^2}^{s_0} d s e^{-{s/T^2}} \rho (s)}
\end{eqnarray}
with $\rho'(s)=s\rho(s)$ except that $\rho'(s)$ does not contain
the last term in $\rho(s)$.

\begin{figure}
\scalebox{0.5}{\includegraphics{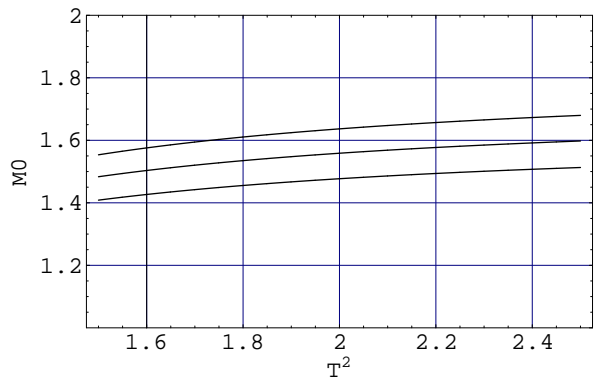}}
\vspace{0.5cm}\caption{The variation of the $I=0$ pentaquark state
mass $M_0$ with the Borel parameter $T$ (in unit of GeV) and the
continuum threshold $s_0$. From bottom to top the curves
correspond to $s_0=3.61, 4.0, 4.41$ GeV$^2$ respectively. }
\label{zhuprl}
\end{figure}

The variation of $M_0$ with the Borel parameter $T$ and the
continuum threshold $s_0$ is shown in Figure (\ref{zhuprl}), which
contributes to the errors of the extracted value, together with
the truncation of the operator product expansion, the uncertainty
of vacuum condensate values and the factorization approximation of
the multiquark condensates. In the working interval of the Borel
parameter $M_0$ is reasonably stable with $T$. Numerically,
\begin{equation}
M_0 =(1.56\pm 0.15) \mbox{GeV} ,
\end{equation}
where the central value corresponds to $T=2$ GeV and $s_0=4$
GeV$^2$. An important observation from Ref. \cite{zhu} is that the
low mass of $\Theta^+$ pentaquark could be accommodated in QCD.

Later, several groups used various forms of interpolating currents
to study the mass and parity of pentaquarks
\cite{qsr,mull,hk,kondo}. However, the continuum contribution to
the resulting mass sum rule is always big because of the high
dimension of the interpolating current, which may render QSR
treatment of pentaquarks less convincing than in the case of
conventional hadrons. Moreover, the convergence of the operator
product expansion is not so good in certain cases.

It's a general result that any pentaquark interpolating currents
can always be decomposed into some linear combinations of the
product of a color-singlet baryon interpolating current and a
color-singlet meson interpolating current. Some examples of the
decomposition were given in \cite{kondo}. When we calculate the
correlation function, we will have two types of contributions to
the spectral density. The first one corresponds to the diagonal
case where a pair of baryon and meson propagate freely. The second
case corresponds to the non-diagonal case when the quark lines of
the baryon mix with those from the meson. I.e., the meson and
baryon pair resonates into a pentaquark which then decays into a
pair of meson and baryon. It is pointed out that the first case is
the background contribution only and has nothing to do with the
pentaquarks \cite{kondo}. Hence this kind of contribution should
be removed.

\subsection{Heptaquarks}

Bicudo proposed an interesting alternate scheme for $\Theta^+$
baryon \cite{bicudo}. Using the resonating group method, he found
that the repulsion is increased by Pauli principle whenever the
two interacting hadrons have a common flavor quark. The attraction
is increased by the quark anti-quark annihilation effect if there
is a pair of quark and anti-quark with the same flavor. The exotic
$\Theta^+$ baryon is unbound if its quark content is $uudd\bar s$.
In order to increase the binding, a light pair of quark and
anti-quark has to be added in order to make the system bound.
Since the pion is the lightest meson, Bicudo proposed that
$\Theta^+$ is probably a $K$$\pi$$N$ molecule with binding energy
of 30 MeV \cite{bicudo}. The decay width has not been calculated
yet in this model. However, one would naively expect such a
loosely bound $K$$\pi$$N$ system is quite broad.

\section{Narrow width puzzle}\label{sec5}

According to these positive-result experiments, both $\Theta^+$
and $\Xi^{--}_5$ are very narrow states. In fact, the $\Theta^+$
pentaquark is so narrow that most of the experiments can only set
an upper bound around $20$ MeV. Recent analysis of kaon nucleon
scattering data indicates that the width of the $\Theta^+$
pentaquark is less than several MeV \cite{nuss}, otherwise they
should have shown up in these old experiments.

Experience with conventional excited hadrons shows that their
widths are around one hundred MeV or even bigger if they lie 100
MeV above threshold and decay through S-wave or P-wave. For
comparison, the $S=-1$ hyperon $\Lambda (1520)$ $D_{03}$ state is
in the same mass region as the $\Theta^+$ pentaquark. Its angular
momentum and parity is $J^P={3\over 2}^-$. Its dominant two-body
decay is of D-wave with final states $N\bar K$, $\Sigma \pi$. With
a smaller phase space and higher partial wave, the width of
$\Lambda (1520)$ is $15.6\pm 1.0$ MeV \cite{pdg}. In contrast,
$\Theta^+$ lies above 100 MeV $K N$ threshold and decays through
either S-wave or P-wave with a total width less than several MeV,
corresponding to negative or positive parity respectively. If
these states are further established and confirmed to have such a
narrow width, the most challenging issue is to understand their
extremely narrow width in a natural way. Is there a mysterious
selection rule which is absent from the conventional hadron
interaction?

In the case of pentaquark decays, if symmetry and kinematics
allow, the most efficient decay mechanism is for the four quarks
and anti-quark to regroup with each other int o a three-quark
baryon and a meson. This is in contrast to the $^3P_0$ decay
models for the ordinary hadrons. This regrouping is coined as the
"fall-apart" mechanism in Refs. \cite{maltman,close,lee,zhanga}.

Recently there have been several attempts to explain the narrow
width of the $\Theta^+$ pentaquark \cite{carl,width2,buccella}.
Carlson et al. constructed a special pentaquark wave function
which is totally symmetric in the flavor-spin part and
anti-symmetric in the color-orbital part in Ref. \cite{carl}. With
this wave function they found that the overlap probability between
the pentaquark and the nucleon kaon system is ${5\over 96}$.
Taking into account of the orbital wave function in JW's diquark
model further reduces the overlap probability to ${5\over 596}$
\cite{carl}. The small overlap probability might be responsible
for the narrow width of pentaquarks.

Karliner and Lipkin proposed that there might exist two nearly
generate pentaquarks \cite{width2}. Both of them decay into the
kaon and nucleon. Hence these two states mix with each other by
the loop diagram via the decay modes. Diagonalization of the mass
matrix leads to a narrow $\Theta^+$ pentaquark which almost
decouples with the decay mode. The pentaquark with the same
quantum number is very broad with a width of around 100 MeV, which
has escaped the experimental detection so far.

In Ref. \cite{buccella} Buccella and Sorba suggested that the four
quarks are in the $L=1$ state and the anti-quark is in the S-wave
state inside the $\Theta^+$ and $\Xi^{--}_5$ pentaquarks. There
are four anti-symmetric four quark $SU(6)_{FS}$ wave functions
$[4]_{FS}, [31]_{FS}, [22]_{FS}, [211]_{FS}$. The third one
corresponds to Jaffe and Wilzcek's diquark model. The difference
is the four quarks are not completely anti-symmetric in the
diquark model. It was pointed out that the narrow width of
$\Theta^+$ pentaquark may favor the last two four-quark
spin-flavor wave functions \cite{buccella}. The reason is as
follows. When the anti-quark picks up a quark to form a meson, two
of the other three quarks remain in the $SU(6)_{fS}$ totally
anti-symmetric states $[21]_{FS}, [111]_{FS}$.  These
representations are orthogonal to the $SU(6)_{FS}$ totally
symmetric representation $[3]_{FS}$ for the nucleon octet. If
$SU(6)_{FS}$ symmetry is exact, the ${\bar 10}$ pentaquarks will
not decay at all. This selection rule is exact in the symmetry
limit. The narrow widths of the $\Theta^+$ and $\Xi^{--}_5$
pentaquarks come from the $SU(3)_f$ symmetry breaking.

Using the picture of the flux tube model, we proposed that the
$\Theta_5$ pentaquark as the first candidate of the
three-dimensional non-planar hadron with the extremely stable
diamond structure. The up and down quarks lie at the corners of
the diamond while the anti-strange quark sits in the center.
Various un-excited color flux tubes between the five quarks bind
them into a stable and narrow color-singlet. Such a configuration
allows the lowest state having the negative parity naturally. The
decay of the $\Theta_5$ pentaquark into the nucleon and kaon
requires the breakup of the non-planar diamond configuration into
two conventional planar hadrons, which involves some kind of
structural phase transition as in the condensed matter physics.
Despite that it lies above the kaon nucleon threshold, the width
of the $\Theta^+$ pentaquark should be narrow because of the
mismatch between the initial and final state spatial wave
functions. Such a mechanism may be tested by future lattice QCD
simulation using non-planar interpolating currents.

\section{Pentaquark flavor wave functions, masses and magnetic
moments}\label{sec6}

The flavor wave functions of diquarks are listed in Table
\ref{tab-diquark}.
\begin{table}[h]
\begin{center}
\begin{tabular}{ccccc}\hline
($Y_1,I_1$)&($Y_2,I_2$) & ($Y,I$) & $I_3$&Flavor Wave Functions\\
\hline
($\frac{1}{3},\frac{1}{2}$)&($\frac{1}{3},\frac{1}{2}$)&($\frac{2}{3},0$)&0&$[ud]=\frac{1}{\sqrt{2}}(ud-du)$\\
($\frac{1}{3},\frac{1}{2}$)&(-$\frac{2}{3}$,0)&(-$\frac{1}{3},\frac{1}{2}$)&$\frac{1}{2}$&$[us]=\frac{1}{\sqrt{2}}(us-su)$\\
(-$\frac{2}{3}$,0)&($\frac{1}{3},\frac{1}{2}$)&(-$\frac{1}{3},\frac{1}{2}$)&-$\frac{1}{2}$&$[ds]=\frac{1}{\sqrt{2}}(ds-sd)$\\
\hline
\end{tabular}
\end{center}
\caption {Diquark flavor wave functions. $Y$, $I$ and $I_3$ are
hypercharge, isospin and the third component of isospin
respectively. The subscripts 1 and 2 represent two quarks inside
the diquark. }\label{tab-diquark}
\end{table}

The flavor wave functions of ${\bar 10}$ and $8$ pentaquarks are
collected in Table \ref{tab-jaffe-wf}.

\begin{table}[h]
\begin{center}
\begin{tabular}{cc||cc}\hline
($Y,I,I_3$)       &$\bf{\bar{10}}$ &($Y,I,I_3$)&$\bf{8}$\\
\hline
(2,0,0)            &$[ud]^2\bar{s}$ &$-$&$-$\\
(1,$\frac{1}{2}$,$\frac{1}{2}$)&$\sqrt{\frac{2}{3}}[ud][us]_+\bar{s}+\sqrt\frac{1}{3}[ud]^2\bar{d}$
     &(1,$\frac{1}{2}$,$\frac{1}{2}$)&$\sqrt\frac{1}{3}[ud][us]_+\bar{s}-\sqrt{\frac{2}{3}}[ud]^2\bar{d}$\\
(1,$\frac{1}{2}$,-$\frac{1}{2}$)&$\sqrt{\frac{2}{3}}[ud][ds]_+\bar{s}+\sqrt\frac{1}{3}[ud]^2\bar{u}$
    &(1,$\frac{1}{2}$,$\frac{1}{2}$)&$\sqrt\frac{1}{3}[ud][ds]_+\bar{s}-\sqrt{\frac{2}{3}}[ud]^2\bar{u}$\\
(0,1,1)
&$\sqrt{\frac{2}{3}}[ud][us]_+\bar{d}+\sqrt\frac{1}{3}[us]^2\bar{s}$
     &(0,1,1)&$\sqrt{\frac{1}{3}}[ud][us]_+\bar{d}-\sqrt\frac{2}{3}[us]^2\bar{s}$\\
(0,1,0) &$\sqrt{\frac{1}{3}}([ud][ds]_+\bar{d}+[ud][us]_+\bar{u}$
     &(0,1,0)&$\sqrt{\frac{1}{6}}([ud][ds]_+\bar{d}+[ud][us]_+\bar{u})$\\
  &$+[us][ds]_+\bar{s})$&&$-\sqrt{\frac{2}{3}}[us][ds]_+\bar{s}$\\
(0,1,-1)
&$\sqrt{\frac{2}{3}}[ud][ds]_+\bar{u}+\sqrt\frac{1}{3}[ds]^2\bar{s}$
     &(0,1,-1)&$\sqrt{\frac{1}{3}}[ud][ds]_+\bar{u}-\sqrt\frac{2}{3}[ds]^2\bar{s}$\\
(-1,$\frac{3}{2}$,$\frac{3}{2}$)&$[us]^2\bar{d}$ &$-$&$-$\\
(-1,$\frac{3}{2}$,$\frac{1}{2}$)&$\sqrt{\frac{2}{3}}[us][ds]_+\bar{d}+\sqrt\frac{1}{3}[us]^2\bar{u}$
     &(-1,$\frac{1}{2}$,$\frac{1}{2}$)&$\sqrt{\frac{1}{3}}[us][ds]_+\bar{d}-\sqrt\frac{2}{3}[us]^2\bar{u}$\\
(-1,$\frac{3}{2}$,-$\frac{1}{2}$&$\sqrt{\frac{2}{3}}[ds][us]_+\bar{u}+\sqrt\frac{1}{3}[ds]^2\bar{d}$
     &(-1,$\frac{1}{2}$,-$\frac{1}{2}$)&$\sqrt{\frac{1}{3}}[ds][us]_+\bar{u}-\sqrt\frac{2}{3}[ds]^2\bar{d}$\\
(-1,$\frac{3}{2}$,-$\frac{3}{2}$)&$[ds]^2\bar{u}$ &$-$&$-$\\
$-$&$-$&(0,0,0)&$\sqrt\frac{1}{2}([ud][ds]_+\bar{d}-[ud][us]_+\bar{u})$\\
 \hline
\end{tabular}
\end{center}
\caption{Flavor wave functions in Jaffe and Wilczek's model
\cite{jaffe}.
$[q_1q_2][q_3q_4]_+=\sqrt{\frac{1}{2}}([q_1q_2][q_3q_4]+[q_3q_4][q_1q_2])$
or $[q_1q_2]^2=[q_1q_2][q_1q_2]$ is the diquark-diquark part.
}\label{tab-jaffe-wf}
\end{table}

Within the diquark model \cite{jaffe}, the strange quark mass
explicitly breaks $SU(3)_F$ symmetry. The [ud] diquark is more
tightly bound than [us] and [ds]. The energy difference can be
related to the $\Sigma$-$\Lambda$ mass splitting. Thus, every
strange quark in the pentaquark contributes $\alpha \equiv \frac
{3}{4}(M_\Sigma - M_\Lambda) \approx 60$ MeV arising from [ud] and
[us], [ds] binding energy difference. The Hamiltonian in JW's
model reads
\begin{equation}
\ H_{s}= M_{0} +(n_{s}+n_{\bar{s}})m_{s}+n_{s} \alpha \label{eq1}
\end{equation}
where $M_{0}$ is the pentaquark mass in the $SU(3)_F$ symmetry
limit. The last two terms are from $SU(3)_F$ symmetry breaking
with $m_s\approx100$ MeV. $M_{0}$ has the form
\begin{equation}
M_{0}=2m_{di}+m_{\bar q}+\delta M_{l} \label{eq2}
\end{equation}
where $m_{di}$ is the [ud] diquark mass, $m_{\bar{q}}$ is the
anti-quark mass and $\delta M_{l}$ is the orbital excitation
energy. We use $\delta M_{l}=0$ for $l=0$ and $\delta M_{l}\approx
230$ MeV for $l=1$, $m_{di}=420$ MeV, and $m_{\bar q}=360$ MeV.
Thus we can use Eq.(\ref{eq1}) to compute pentaquark masses. For
example, we have
\begin{equation}
M_{\Theta^+}\approx 1530 \mbox{MeV}.
\end{equation}
In Table \ref{tab-jaffe-neg} we have listed the flavor wave
functions and masses of $J^P={1\over 2}^-$ pentaquarks in JW's
model.
\begin{table}[h]
\begin{center}
\begin{tabular}{c|cc|c|c}\hline
&($Y,I$)               &   $I_3$       &Flavor wave functions&Masses  \\
\hline
$p_8$&(1,$\frac{1}{2}$)&$\frac{1}{2}$ &$[su][ud]_-\bar{s}$ &1460\\
$n_8$&         &-$\frac{1}{2}$ &$[ds][ud]_-\bar{s}$&1460 \\
$\Sigma_{8}^+$&(0,1)          &1 &$[su][ud]_-\bar{d}$ &1360\\
$\Sigma_{8}^0$&                &0  &$\frac{1}{\sqrt{2}}$($[su][ud]_-\bar{u}+[ds][ud]_-\bar{d}$)&1360 \\
$\Sigma_{8}^-$&                &-1          &$[ds][ud]_-\bar{u}$&1360 \\
$\Lambda_8$&(0,0)             & 0 &$\frac{1}{\sqrt6}([ud][su]_-\bar{u}+[ds][ud]_-\bar{d}-2[su][ds]_-\bar{s})$&1533\\
$\Xi_{8}^0$&  (-1,$\frac{1}{2}$)  & $ \frac{1}{2}$               &$[ds][su]_-\bar{d}$&1520 \\
$\Xi_{8}^-$&                 &-$\frac{1}{2}$ &$[ds][su]_-\bar{u}$ &1520 \\
\hline
$\Lambda_1$&(0,0)&0                &$\frac{1}{\sqrt3}([ud][su]_-\bar{u}+[ds][ud]_-\bar{d}+[su][ds]_-\bar{s})$&1447\\
\hline
\end{tabular}
\end{center}
\caption{Flavor wave functions and masses of the $\frac{1}{2}^-$
pentaquark octet and singlet. Where
$[q_1q_2][q_3q_4]_-=\sqrt{\frac{1}{2}}([q_1q_2][q_3q_4]-[q_3q_4][q_1q_2])$.
}\label{tab-jaffe-neg}
\end{table}

The baryon magnetic moment is another fundamental observable as
its mass, which encodes information of the underlying quark
structure and dynamics. Different models generally yield different
values, which provides a very good way to distinguish them. This
in turn will affect the photo-production and electro-production
cross sections of pentaquarks.

The magnetic moment of a compound system is the sum of the
magnetic moments of its constituents including spin and orbital
contributions,
\begin{equation}
\overrightarrow{\mu}=\sum\limits_{i}\overrightarrow{\mu_i}
=\sum\limits_{i}(g_{i}\overrightarrow{s_i}+\overrightarrow{l_i})
\mu_i,
\end{equation}
where $g_{i}$ is the $g$-factor of $i$-th constituent and $\mu_i$
is the magneton of the $i$-th constituent
\begin{equation}
\mu_i = \frac{e_i}{2m_i}\; .
\end{equation}
Using the diquark flavor wave functions Table \ref{tab-diquark}
and pentaquark flavor wave functions in Table \ref{tab-jaffe-wf}
and \ref{tab-jaffe-neg}, one can easily derive pentaquark magnetic
moments in this model \cite{liu-mm}. The results are listed in
Table \ref{tab-jaffe-mm1} and \ref{tab-jaffe-mm2}.
\begin{table}[h]
\begin{center}
\begin{tabular}{c|c|c||c|c|c}\hline
 State    & \multicolumn{2}{|c||}{$\bf{\bar{10}}$}&State&\multicolumn{2}{|c}{\bf{8}}\\
 \hline
 ($Y,I,I_3$)                  & set I &set II &($Y,I,I_3$)        & set I &set II\\
 \hline
(2,0,0)                     &0.08 &0.29 &$-$&$-$&$-$  \\
(1,$\frac{1}{2}$,$\frac{1}{2}$) &0.037&0.22&(1,$\frac{1}{2}$,$\frac{1}{2}$)&0.018&0.21\\
(1,$\frac{1}{2}$,-$\frac{1}{2}$)&0.12 &0.22&(1,$\frac{1}{2}$,-$\frac{1}{2}$)&0.50 &0.65\\
(0,1,1)                     &-0.009&0.14&(0,1,1)&0.007&0.14\\
(0,1,0)                     &0.06 &0.12&(0,1,0)&-0.13&-0.13\\
(0,1,-1)                    &0.13 &0.09&(0,1,-1)&-0.27&-0.41\\
(-1,$\frac{3}{2}$,$\frac{3}{2}$)&-0.06&0.06 &$-$&$-$&$-$\\
(-1,$\frac{3}{2}$,$\frac{1}{2}$)&0     &0   &(-1,$\frac{1}{2}$,$\frac{1}{2}$) &0.41 &0.46\\
(-1,$\frac{3}{2}$,-$\frac{1}{2}$)&0.06&-0.06&(-1,$\frac{1}{2}$,-$\frac{1}{2}$)&-0.35&-0.52\\
(-1,$\frac{3}{2}$,-$\frac{3}{2}$)&0.12&-0.12&$-$&$-$&$-$\\
$-$&$-$&$-$&(0,0,0) &0.25 &0.37\\
\hline
\end{tabular}
\end{center}
\caption{Numerical results of the magnetic moments of ${\bf \bar{
10}}$ and {\bf 8} pentaquarks in unit of $\mu_N$ in Jaffe and
Wilczek's model \cite{jaffe}. For set I we use $m_{ud}=720$ MeV,
$m_{us}=m_{ds}=900$ MeV from Ref. \cite{lipkin} and for set II we
use $m_{ud}=420$ MeV, $m_{us}=m_{ds}=600$ MeV from Ref.
\cite{jaffe}. }\label{tab-jaffe-mm1}
\end{table}

\begin{table}[h]
\begin{center}
\begin{tabular}{c|cc|c|c}\hline
&($Y,I$)               &   $I_3$       &Magnetic moments & Numerical results ($\mu_N$)\\
\hline
$p_8$&(1,$\frac{1}{2}$)&$\frac{1}{2}$&$\frac{e_0}{6m_s}$ &0.63\\
$n_8$&               &-$\frac{1}{2}$&$\frac{e_0}{6m_s}$ &0.63 \\
$\Sigma_{8}^+$&(0,1)          &1 &$\frac{e_0}{6m_d}$ &0.87\\
$\Sigma_{8}^0$&                &0                &$\frac{1}{6}(-\frac{e_0}{m_u}+\frac{e_0}{2m_d})$&-0.43\\
$\Sigma_{8}^-$&                &-1          &$-\frac{e_0}{3m_u}$&-1.74\\
$\Lambda_8$  &(0,0)& 0 &$\frac{1}{18}$(-$\frac{e_0}{m_u}$+$\frac{e_0}{2m_d}$+$\frac{2e_0}{m_s}$)&0.27\\
$\Xi_{8}^0$& (-1,$\frac{1}{2}$)  & $ \frac{1}{2}$               &$\frac{e_0}{6m_d}$ &0.87\\
$\Xi_{8}^-$&               &-$\frac{1}{2}$ &-$\frac{e_0}{3m_u}$ &-1.74\\
\hline
$\Lambda_1$&(0,0)&0                &$\frac{1}{9}(-\frac{e_0}{m_u}+\frac{e_0}{2m_d}+\frac{e_0}{2m_s})$&-0.08\\
\hline
\end{tabular}
\end{center}
\caption{Expressions and numerical results of the  magnetic
moments of the pentaquark octet and singlet, where $e_0$ is the
charge unit. } \label{tab-jaffe-mm2}
\end{table}

For the sake of comparison, $J^P={1\over 2}^+$ pentaquark magnetic
moments and flavor wave functions in Karliner and Lipkin's
triquark model are also listed in Table \ref{tab-kl-mm1} and
\ref{tab-kl-mm2}. Magnetic moments of $J^P={3\over 2}^+$ can be
found in Refs. \cite{li-mm}.
\begin{table}
\begin{center}
\begin{tabular}{ccc} \hline
($Y,I,I_3$)        &  Flavor wave functions     &Magnetic moments
\\ \hline
(2,0,0)                              &$[ud]\{ud\bar{s}\}$                                                          &   0.19   \\
(1,$\frac{1}{2}$,$\frac{1}{2}$)          &$\sqrt{\frac{1}{3}}([ud]\{ud\bar{d}\}+[ud]\{us\bar{s}\}+[us]\{ud\bar{s}\})$  &   0.17   \\
(1,$\frac{1}{2}$,-$\frac{1}{2}$)       &$\sqrt{\frac{1}{3}}([ud]\{ud\bar{u}\}+[ud]\{ds\bar{s}\}+[ds]\{ud\bar{s}\})$  &   -0.006  \\
(0,1,1)                   &$\sqrt{\frac{1}{3}}([us]\{ud\bar{d}\}+[us]\{us\bar{s}\}+[ud]\{us\bar{d}\})$             &   0.15   \\
(0,1,0)                   &$\sqrt{\frac{1}{6}}([us]\{ud\bar{u}\}+[us]\{ds\bar{s}\}+[ds]\{ud\bar{d}\}$              &   -0.03  \\
                            &$+[ds]\{us\bar{s}\}+[ud]\{us\bar{u}\}+[ud]\{ds\bar{d}\})$&\\
(0,1,-1)                &$\sqrt{\frac{1}{3}}([ds]\{ud\bar{u}\}+[ds]\{ds\bar{s}\}+[ud]\{ds\bar{u}\})$             &   -0.21  \\
(-1,$\frac{3}{2}$,$\frac{3}{2}$)       &$[us]\{us\bar{d}\}$                                                          &   0.13   \\
(-1,$\frac{3}{2}$,$\frac{1}{2}$)       &$\sqrt{\frac{1}{3}}([us]\{us\bar{u}\}+[us]\{ds\bar{d}\}+[ds]\{us\bar{d}\})$  &   -0.054  \\
(-1,$\frac{3}{2}$,-$\frac{1}{2}$)    &$\sqrt{\frac{1}{3}}([ds]\{ds\bar{d}\}+[ds]\{us\bar{u}\}+[us]\{ds\bar{u}\})$  &   -0.24  \\
(-1,$\frac{3}{2}$,-$\frac{3}{2}$)    &$[ds]\{ds\bar{u}\}$                                                            &   -0.43  \\
\hline
\end{tabular}
\caption{Wave functions and numerical results of the magnetic
moments of ${\bf \bar{10}}$ pentaquarks in unit of $\mu_N$ in
Karliner and Lipkin's model \cite{lipkin}. $Y$, $I$ and $I_3$ are
hypercharge, isospin and the third component of isospin
respectively. $\{q_1q_2\bar{q_3}\}\equiv[q_1q_2]\bar{q_3}$ is the
triquark's flavor wave function.}\label{tab-kl-mm1}
\end{center}
\end{table}

\begin{table}
\begin{center}
\begin{tabular}{ccc} \hline
($Y,I,I_3$)        &  Flavor wave functions     &Magnetic moments
\\ \hline
(1,$\frac{1}{2}$,$\frac{1}{2}$)      &$\sqrt{\frac{1}{6}}([ud]\{ud\bar{d}\}+[ud]\{us\bar{s}\})-\sqrt{\frac{2}{3}}[us]\{ud\bar{s}\}$  &   0.16    \\
(1,$\frac{1}{2}$,-$\frac{1}{2}$)     &$\sqrt{\frac{1}{6}}([ud]\{ud\bar{u}\}+[ud]\{ds\bar{s}\})-\sqrt{\frac{2}{3}}[ds]\{ud\bar{s}\}$  &   -0.14   \\
(0,1,1)                              &$\sqrt{\frac{1}{6}}([us]\{ud\bar{d}\}+[us]\{us\bar{s}\})-\sqrt{\frac{2}{3}}[ud]\{us\bar{d}\}$  &   0.16    \\
(0,1,0)                              &$\sqrt{\frac{1}{12}}([us]\{ud\bar{u}\}+[us]\{ds\bar{s}\}+[ds]\{ud\bar{d}\}$                    &   0.04    \\
                                       &$+[ds]\{us\bar{s}\})-\sqrt{\frac{1}{3}}([ud]\{us\bar{u}\}+[ud]\{ds\bar{d}\})$&\\
(0,1,-1)                             &$\sqrt{\frac{1}{6}}([ds]\{ud\bar{u}\}+[ds]\{ds\bar{s}\})-\sqrt{\frac{2}{3}}[ud]\{ds\bar{u}\}$  &   -0.07   \\
(-1,$\frac{1}{2}$,$\frac{1}{2}$)     &$\sqrt{\frac{1}{6}}([us]\{us\bar{u}\}+[us]\{ds\bar{d}\})-\sqrt{\frac{2}{3}}[ds]\{us\bar{d}\}$  &   -0.17   \\
(-1,$\frac{1}{2}$,-$\frac{1}{2}$)    &$\sqrt{\frac{1}{6}}([ds]\{ds\bar{d}\}+[ds]\{us\bar{u}\})-\sqrt{\frac{2}{3}}[us]\{ds\bar{u}\}$  &   -0.12   \\
(0,0,0)                              &$\frac{1}{2}([us]\{ud\bar{u}\}+[us]\{ds\bar{s}\}-[ds]\{ud\bar{d}\}-[ds]\{us\bar{s}\})$         &   -0.10   \\
\hline
\end{tabular}
\caption{Wave functions and numerical results of the magnetic
moments of ${\bf {8_f}}$ pentaquarks in unit of $\mu_N$ in
Karliner and Lipkin's model \cite{lipkin}. $Y$, $I$ and $I_3$ are
hypercharge, isospin and the third component of isospin
respectively.}\label{tab-kl-mm2}
\end{center}
\end{table}

The magnetic moments of $J=\frac{1}{2}$ pentaquarks have been
discussed by several groups recently. Using Jaffe and Wilczek's
scalar diquark picture of $\Theta^+$, Nam, Hosaka and Kim
\cite{hosaka1} considered the photoproduction of $\Theta^+$
pentaquark from the neutron and estimated the anomalous magnetic
moment to be $- 0.7 $ (positive parity) and $- 0.2$ (negative
parity) (in units of $\Theta^+$ magneton ${e_0\over
2m_{\Theta^+}}$).

In Ref. \cite{zhao} a quark model calculation is also performed
using JW's  diquark picture, where Zhao got
$\mu_{\Theta^+}=0.13{e_0\over 2m_{\Theta^+}}$ for positive parity
$\Theta^+$. In the case of negative parity, he treated the
pentaquark as the sum of $(u\bar s)$ and $(udd)$ clusters and got
$\mu_{\Theta^+}={e_0\over 6m_s}$.

The magnetic moment of the $\Theta^+$ pentaquark is also
calculated using the method of light cone QCD sum rules. In Ref.
\cite{huang} the authors arrived at $\mu_{\Theta^+}=(0.12\pm 0.06)
\mu_N$.

Within the chiral soliton model, Kim and Praszalowicz \cite{mag}
derived relations for the $\bf{\bar{10}_f}$ magnetic moments and
found that the magnetic moment of the $\Theta^+$ pentaquark is
between $-(0.3\sim 0.4) \mu_N$ in the chiral limit.

Using the non-clustered quark model, the $\Theta^+$ pentaquark is
found to be $0.38 \mu_N, 0.09 \mu_N, 1.05 \mu_N$ for $J^P={1\over
2}^-, {1\over 2}^+, {3\over 2}^+$ respectively in Ref.
\cite{bijker}. For $\Xi^{--}, \Xi^+$, the numerical values are
$-0.44 \mu_N, 0.09 \mu_N, -2.64 \mu_N$ and $0.5 \mu_N, 0.09 \mu_N,
1.59 \mu_N$ respectively for $J^P={1\over 2}^-, {1\over 2}^+,
{3\over 2}^+$.

We have collected the numerical results for $J=\frac{1}{2}$
$\bf{\bar {10}}$ members $\Theta^+$, $\Xi_5^{--}$, $\Xi_5^+$ and
$J=\frac{3}{2}$ $\bf{\bar {10}}$ members $\Theta^{\ast +}$,
$\Xi_5^{\ast --}$, $\Xi_5^{\ast +}$ in Table \ref{tab-summary}.
These states lie on the corners of the anti-decuplet triangle and
have no mixing with octet pentaquarks. Hence their interpretation
and identification should be relatively clean, at least
theoretically.
\begin{table}
\caption{Comparison of magnetic moments of $\Theta^+$,
$\Xi^{--}_5$, $\Xi^{+}_5$ and $\Theta^{\ast +}$, $\Xi^{\ast
--}_5$, $\Xi^{\ast +}_5$ in different pentaquark models in
literature. The numbers are in unit of
$\mu_N$.}\label{tab-summary}
\begin{center}
\begin{tabular}{c|c|c|c|c|c|c}
\hline

                            &  \multicolumn{3}{c}{$J=1/2$}\vline   &  \multicolumn{3}{c}{$J=3/2$}       \\
\hline

                            &   $\Theta^+$   & $\Xi^{--}_5$ & $\Xi^{+}_5$    &  $\Theta^{\ast +}$ & $\Xi^{\ast --}_5$ & $\Xi^{\ast +}_5$  \\

\hline
Ref. \cite{mag}             & $-(0.3\sim 0.4)$  &    $-0.4 $        &      $0.2$         &     -       &       -      &       -      \\
\hline
Ref. \cite{hosaka}          & $0.2\sim 0.5$  &    -         &      -         &     -       &       -      &       -      \\
\hline
Ref. \cite{zhao}            & $0.08\sim 0.6$ &    -         &      -         &     -       &       -      &       -      \\
\hline
Ref. \cite{bijker}           & $0.09\sim 0.38$ &    $0.09\sim -0.44$     &  $0.09\sim 0.50$        &     1.05       &       -2.64      &     1.59      \\
\hline
Ref. \cite{huang}           & $0.12\pm 0.06$ &    -         &      -         &     -       &       -      &       -      \\
\hline
Ref. \cite{li-mm} (JW's model)   &      0.08    &     0.12     &    -0.06       &    1.01     &    -2.43     &    1.22       \\
\hline
Ref. \cite{li-mm} (KL's model) &      0.19    &     -0.43    &     0.13       &    0.84     &    -1.20     &    0.89       \\
\hline
\end{tabular}
\end{center}
\end{table}

\section{Heavy flavored pentaquarks}

It is straightforward to extend the same formalism to discuss the
heavy pentaquarks containing an anti-charm or anti-bottom quark.
One gets one $SU(3)_f$ anti-sextet in the framework of the diquark
model and both a sextet and triplet in KL's models with either
$J^P={1\over 2}^+$ or ${3\over 2}^+$.

As in the case of conventional heavy hadrons, the presence of the
heavy anti-quark makes the treatment of the system simpler. If the
light quarks are really strongly correlated as proposed in the
clustered quark models, the heavy pentaquark system is the ideal
place to study this kind of correlation without the additional
complication due to the extra light anti-quark.

\begin{figure}[h]
\begin{center}
\begin{picture}(200,130)

\thicklines \put(50,22){\line(3,5){50}}
\put(100,105){\line(3,-5){50}} \put(50,22){\line(1,0){100}}
\put(100,22){\line(3,5){25}}\put(100,22){\line(-3,5){25}}\put(75,64){\line(1,0){50}}
\put(50,22){\circle*{4}}\put(100,22){\circle*{4}}\put(150,22){\circle*{4}}
\put(75,64){\circle*{4}}\put(125,64){\circle*{4}}\put(100,105){\circle*{4}}

\thinlines
\put(30,50){\line(1,0){140}}\put(100,2){\line(0,1){120}}

\put(105,105){$\Theta_c^0$,$\Theta_b^+$}
\put(130,64){$\Sigma_{5c}^0$,$\Sigma_{5b}^+$}
\put(35,64){$\Sigma_{5c}^-$,$\Sigma_{5b}^0$}
\put(155,10){$\Xi_{5c}^0$,$\Xi_{5b}^+$}
\put(103,10){$\Xi_{5c}^-$,$\Xi_{5b}^0$}
\put(33,10){$\Xi_{5c}^-$,$\Xi_{5b}^0$}
\end{picture}
\end{center}
\caption{The six members of the $SU(3)$ flavor
anti-sextet.}\label{fig-heavy}
\end{figure}
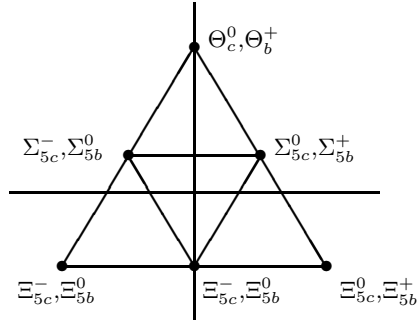

The heavy anti-quark is a $SU(3)_f$ flavor singlet. Hence the
heavy pentaquarks containing an anti-charm or anti-bottom form a
$SU(3)$ flavor anti-sextet in the diquark model as shown in Figure
\ref{fig-heavy}. The flavor wave functions of all members of ${\bf
{\bar 6_f}}$ are listed in Table \ref{heavy-jaffe-wf}.

\begin{table}[h]
\begin{center}
\begin{tabular}{c|c|c}\hline
pentaquarks                    &    $(Y,I,I_3)$                              & Flavor wave functions($\bar Q=\bar c$ or $\bar b$) \\
\hline
$\Theta^0_c$,$\Theta^+_b$      &   ($\frac{4}{3}$,0,0)                       &     $[ud]^2 \bar{Q}$      \\
$\Sigma^0_{5c}$,$\Sigma^+_{5b}$& ($\frac{1}{3}$,$\frac{1}{2}$,$\frac{1}{2}$) &     $[ud][us]_+\bar{Q}$   \\
$\Sigma^-_{5c}$,$\Sigma^0_{5b}$& ($\frac{1}{3}$,$\frac{1}{2}$,-$\frac{1}{2}$)&     $[ud][ds]_+\bar{Q}$   \\
$\Xi^0_{5c}$,$\Xi^+_{5b}$      & (-$\frac{2}{3}$,1,1)                        &     $[us]^2\bar{Q}$       \\
$\Xi^-_{5c}$,$\Xi^0_{5b}$      &(-$\frac{2}{3}$,1,)                          &     $[us][ds]_+\bar{Q}$   \\
$\Xi^{--}_{5c}$,$\Xi^-_{5b}$   &(-$\frac{2}{3}$,1,-1)                        &     $[ds]^2\bar{Q}$       \\
 \hline
\end{tabular}
\end{center}
\caption{Flavor wave functions in Jaffe and Wilczek's model
\cite{jaffe}.
$[q_1q_2][q_3q_4]_+=\sqrt{\frac{1}{2}}([q_1q_2][q_3q_4]+[q_3q_4][q_1q_2])$
or $[q_1q_2]^2=[q_1q_2][q_1q_2]$ is the diquark-diquark part.
}\label{heavy-jaffe-wf}
\end{table}

Jaffe and Wilczek estimated $\Theta^0_c$ and $\Theta^+_b$ masses
by replacing the $\bar s$ in the $\Theta^+$ with $\bar c$ and
$\bar b$ respectively. The cost of this replacement is roughly the
mass difference between $\Lambda_c$ and $\Lambda$ baryons. The
$[ud]$ diquark in the $\Lambda_c$ and $\Lambda$ experiences nearly
the same environment as in $\Theta^0_c$ and $\Theta^+$
pentaquarks, especially when the $[ud]$ diquark is viewed as a
tightly bound entity. Thus they got
\begin{eqnarray}\nonumber
M(\Theta^0_c)=M(\Theta^+)+[M(\Lambda_c)-M(\Lambda)]\\\nonumber
M(\Theta^+_b)=M(\Theta^+)+[M(\Lambda^+_b)-M(\Lambda)].\nonumber
\end{eqnarray}

Their formalism can be extended to estimate the masses of all the
other members of ${\bf {\bar 6_f}}$. The heavy pentaquarks in the
same isospin multiplet are degenerate.
\begin{eqnarray}\nonumber
M(\Sigma^0_{5c})=M(N^+_s)+[M(\Lambda_c)-M(\Lambda)]\\\nonumber
M(\Sigma^+_{5b})=M(N^+_s)+[M(\Lambda^+_b)-M(\Lambda)].\nonumber
\end{eqnarray}
Here $N^+_s$ has a quark content $|[ud][us]_+ \bar s\rangle$. Its
mass was estimated to be around $1700$ MeV \cite{jaffe}. For the
heavy pentaquark containing two strange quarks,
\begin{eqnarray}\nonumber
M(\Xi^0_{5c})=M(\Xi^{--}_5)+[M(\Xi_c)-M(\Sigma)]\\\nonumber
M(\Xi^+_{5b})=M(\Xi^{--}_5)+[M(\Xi_b)-M(\Sigma)],\nonumber
\end{eqnarray}
where $\Xi^{--}_5 $ mass was taken from NA49 experiment
\cite{na49}. The heavy pentaquark masses are collected in Table
\ref{heavy-jaffe-mass} \cite{huang-heavy}.

\begin{table}[h]
\begin{center}
\begin{tabular}{c|c|c|c}\hline
$\bar Q=\bar c$               &   Masses  &     $\bar Q=\bar b$         &  Mass    \\
\hline
$\Theta^0_c$                               & $2710$MeV &          $\Theta^+_b$                    & $6050$MeV  \\
$\Sigma^-_{5c}$,$\Sigma^0_{5c}$            & $2870$MeV &    $\Sigma^0_{5b}$,$\Sigma^+_{5b}$       & $6210$MeV  \\
$\Xi^{--}_{5c}$,$\Xi^-_{5c}$,$\Xi^0_{5c}$  & $3135$MeV &  $\Xi^-_{5b}$,$\Xi^0_{5b}$,$\Xi^+_{5b}$  & $6475$MeV  \\
\hline
\end{tabular}
\end{center}
\caption{Masses of all members of ${\bf {\bar 6_f}}$ in Jaffe and
Wilczek's model. }\label{heavy-jaffe-mass}
\end{table}

It is interesting to note that the resulting $\Theta_c$ mass 2910
MeV is 389 MeV smaller than the experimental value 3099 MeV from
the H1 experiment. If the diquark model is a good framework for
pentaquarks, there is one possible explanation of this big
discrepancy. The anti-quark mass dependent interaction in
$\Theta^+$ between the diquark pair and $\bar s$ is much more
attractive than that in $\Theta_c$ between the same diquark pair
and $\bar c$ since the charm quark is much heavier than the
strange quark. With a bigger value of the diquark mass and the
inclusion of this attractive interaction between the diquark pair
and the anti-quark, one may be able to accommodate both $\Theta^+$
and $\Theta_c$ mass simultaneously \cite{haiyang}.

In Table \ref{heavy-jaffe-mag} we have collected the magnetic
moments of the heavy pentaquarks in the diquark model.

\begin{table}[h]
\begin{center}
\begin{tabular}{c|c|c|c|c}\hline
&\multicolumn{2}{c}{Q=c}\vline&\multicolumn{2}{c}{Q=b}\\
\cline{2-5}
  ($Y,I,I_3$) &$J^P={1\over 2}^+$&$J^P={3\over 2}^+$&$J^P={1\over 2}^+$&$J^P={3\over 2}^+$\\
\hline
($\frac{4}{3}$,0,0)                        &0.62&0.38&0.48&0.81\\
($\frac{1}{3}$,$\frac{1}{2}$,$\frac{1}{2}$)&0.56&0.29&0.41&0.71\\
($\frac{1}{3}$,$\frac{1}{2}$,-$\frac{1}{2}$)&0.13&-0.36&-0.015&0.071\\
(-$\frac{2}{3}$,1,1)                        &0.47&0.16&0.33&0.58\\
(-$\frac{2}{3}$,1,0)                       &-0.052&-0.63&-0.19&-0.20\\
(-$\frac{2}{3}$,1,-1)                      &-0.57&-1.41&-0.72&-0.98\\
 \hline
\end{tabular}
\end{center}
\caption{Numerical results of the $J^P=\frac{1}{2}^+$ and
$J^P=\frac{3}{2}^+$ heavy pentaquark magnetic moments in Jaffe and
Wilczek's model with $m_{ud}=420$ MeV, $m_{us}=m_{ds}=600$ MeV
from Ref. \cite{jaffe}. }\label{heavy-jaffe-mag}
\end{table}

In Karliner and Lipkin's model, the direct product of the ${\bf
\bar 3_f}$ of diquark and the ${\bf \bar 3_f}$ of triquark leads
to ${\bf\bar 6_f}$ and ${\bf 3_f}$ pentaquarks. There is one
orbital angular momentum $L=1$ between the diquark and the
triquark. The resulting $J^P$ of the pentaquark can be either
${1\over 2}^+$ or ${3\over 2}^+$. We list the flavor wave
functions of these two multiplets in Table \ref{heavy-lipkin-wf1}
and \ref{heavy-lipkin-wf2}.

\begin{table}
\begin{center}
\begin{tabular}{c|c|c} \hline
pentaquarks &($Y,I,I_3$)        &  Flavor wave functions
\\ \hline
$\Theta^0_c$,$\Theta^+_b$      &($\frac{4}{3}$,0,0)                           &$[ud]\{ud\bar{Q}\}$        \\
$\Sigma^0_{5c}$,$\Sigma^+_{5b}$&($\frac{1}{3}$,$\frac{1}{2}$,$\frac{1}{2}$)   &$\sqrt{\frac{1}{2}}([ud]\{us\bar{Q}\}+[us]\{ud\bar{Q}\})$\\
$\Sigma^-_{5c}$,$\Sigma^0_{5b}$&($\frac{1}{3}$,$\frac{1}{2}$,-$\frac{1}{2}$)  &$\sqrt{\frac{1}{2}}([ud]\{ds\bar{Q}\}+[ds]\{ud\bar{Q}\})$\\
$\Xi^0_{5c}$,$\Xi^+_{5b}$      &(-$\frac{2}{3}$,1,1)                          &$[us]\{us\bar{Q}\}$        \\
$\Xi^-_{5c}$,$\Xi^0_{5b}$      &(-$\frac{2}{3}$,1,0)                          &$\sqrt{\frac{1}{2}}([us]\{ds\bar{Q}\}+[ds]\{us\bar{Q}\}$     \\
$\Xi^{--}_{5c}$,$\Xi^-_{5b}$   &(-$\frac{2}{3}$,1,-1)                         &$[ds]\{ds\bar{Q}\}$        \\
\hline
\end{tabular}
\caption{Flavor wave functions of the anti-sextet pentaquarks in
Karliner and Lipkin's model \cite{lipkin}. $Y$, $I$ and $I_3$ are
hypercharge, isospin and the third component of isospin
respectively. $\{q_1q_2\bar{Q}\}\equiv[q_1q_2]\bar{Q}$ is the
triquark's flavor wave function.}\label{heavy-lipkin-wf1}
\end{center}
\end{table}

\begin{table}
\begin{center}
\begin{tabular}{c|c|c} \hline
pentaquarks   &($Y,I,I_3$)        &  Flavor wave functions
\\ \hline
$\Sigma'^0_{5c}$,$\Sigma'^+_{5b}$&($\frac{1}{3}$,$\frac{1}{2}$,$\frac{1}{2}$) &$\sqrt{\frac{1}{2}}([ud]\{us\bar{Q}\}-[us]\{ud\bar{Q}\})$\\
$\Sigma'^-_{5c}$,$\Sigma'^0_{5b}$&($\frac{1}{3}$,$\frac{1}{2}$,-$\frac{1}{2}$)&$\sqrt{\frac{1}{2}}([ud]\{ds\bar{Q}\}-[ds]\{ud\bar{Q}\})$\\
$\Xi'^-_{5c}$,$\Xi'^0_{5b}$      &(-$\frac{2}{3}$,0,0)                        &$\sqrt{\frac{1}{2}}([us]\{ds\bar{Q}\}-[ds]\{us\bar{Q}\}$ \\
\hline
\end{tabular}
\caption{Flavor wave functions of the triplet heavy pentaquarks in
Karliner and Lipkin's model \cite{lipkin}.
}\label{heavy-lipkin-wf2}
\end{center}
\end{table}

Karliner and Lipkin have estimated the masses of $\Theta^+$ and
its heavy flavor analogs \cite{lipkin}. The mass of a pentaquark
comes mainly from the masses of its constituent quarks, the
color-spin hyperfine interaction and the P-wave excitation between
the two clusters. The first two parts can be estimated by
comparing them with a relevant baryon-meson system that have the
same quark contents as this pentaquark. The hyperfine energy
difference between these two systems can be figured out using the
$SU(6)$ color-spin algebra and has been given as
$-\frac{1+\zeta_Q}{12}[M(\Delta)-M(N)]$ by Karliner and Lipkin
\cite{lipkin}, where $\zeta_Q\equiv\frac{m_u}{m_Q}$. The P-wave
excitation energy $\delta E^P$ is approximate to the mass
difference between $D_s(2319)$ and $D^*_s(2112)$. This
approximation is based on the observation that the reduced mass of
the diquark-triquark system is close to that of the $D_s$
\cite{lipkin}.

Following this formalism the masses of other members of ${\bf
{\bar 6_f}}$ and ${\bf 3_f}$ read
\begin{eqnarray}\nonumber
 \lefteqn{M(\Sigma^-_{5c})=M(\Sigma^0_{5c})=M(\Sigma'^-_{5c})=M(\Sigma'^0_{5c})}\\\nonumber
 &=\frac{1}{2}[M(N)+M(D_s)+M(\Sigma)+M(D)]
 \\\nonumber
 &-\frac{1+\zeta_c}{12}[M(\Delta)-M(N)]+\delta
 E^{P}\\\nonumber
 \lefteqn{M(\Sigma^0_{5b})=M(\Sigma^+_{5b})=M(\Sigma'^0_{5b})=M(\Sigma'^+_{5b})}\\\nonumber
 &=\frac{1}{2}[M(N)+M(B_s)+M(\Sigma)+M(B)]
 \\\nonumber
 &-\frac{1+\zeta_b}{12}[M(\Delta)-M(N)]+\delta
 E^{P}\nonumber
\end{eqnarray}
and
\begin{eqnarray}\nonumber
 \lefteqn{M(\Xi^{--}_{5c})=M(\Xi^-_{5c})=M(\Xi'^-_{5c})=M(\Xi^0_{5c})}\\\nonumber
 &=M(\Sigma)+M(D_s)]-\frac{1+\zeta_c}{12}[M(\Delta)-M(N)]+\delta E^{P}\\\nonumber
 \lefteqn{M(\Xi^-_{5b})=M(\Xi^0_{5b})=M(\Xi'^0_{5b})=M(\Xi^+_{5b})}\\\nonumber
 &=M(\Sigma)+M(B_s)]-\frac{1+\zeta_b}{12}[M(\Delta)-M(N)]+\delta E^{P}.\nonumber
\end{eqnarray}
The numerical values of the masses of all members of ${\bf {\bar
6_f}}$ and ${\bf 3_f}$ are presented in Table
\ref{heavy-lipkin-mass}.

\begin{table}[h]
\begin{center}
\begin{tabular}{c|c|c|c}\hline
$\bar Q=\bar c$     &   Mass  &     $\bar Q=\bar b$         &  Mass    \\
\hline
$\Theta^0_c$                     & $2990$MeV &          $\Theta^+_b$                    & $6400$MeV  \\
$\Sigma^-_{5c}(\Sigma'^-_{5c})$, $\Sigma^0_{5c}(\Sigma'^0_{5c})$ &
$3165$MeV & $\Sigma^0_{5b}(\Sigma'^0_{5b})$,
                                               $\Sigma^+_{5b}(\Sigma'^+_{5b})$          & $6570$MeV  \\
$\Xi^{--}_{5c}$,$\Xi^-_{5c}(\Xi'^-_{5c})$,$\Xi^0_{5c}$
                                 & $3340$MeV & $\Xi^-_{5b}$,$\Xi^0_{5b}(\Xi'^0_{5b})$,
                                               $\Xi^+_{5b}$                             & $6740$MeV  \\
\hline
\end{tabular}
\end{center}
\caption{Masses of all members of ${\bf {\bar 6_f}}$ and ${\bf {
3_f}}$ in Karliner and Lipkin's model. }\label{heavy-lipkin-mass}
\end{table}

For comparison, the heavy pentaquark magnetic moments are shown in
Table \ref{heavy-lipkin-mag}.

\begin{table}[h]
\begin{center}
\begin{tabular}{c|c|c|c|c}\hline
pentaquarks  &\multicolumn{2}{c}{Q=c}\vline&\multicolumn{2}{c}{Q=b}\\
\cline{2-5}
   &$J^P={1\over 2}^+$&$J^P={3\over 2}^+$&$J^P={1\over 2}^+$&$J^P={3\over 2}^+$\\
\hline
$\Theta^0_c$,$\Theta^+_b$                                          &-0.031 &1.01  &0.079   &0.96  \\
$\Sigma^0_{5c}$,$\Sigma'^0_{5c}$,$\Sigma^+_{5b}$,$\Sigma'^+_{5b}$  &-0.079 &1.06  &0.030   &1.00  \\
$\Sigma^-_{5c}$,$\Sigma'^-_{5c}$,$\Sigma^0_{5b}$,$\Sigma'^0_{5b}$  &-0.084 &-0.26 &-0.0028 &-0.35 \\
$\Xi^0_{5c}$,$\Xi^+_{5b}$                                          &-0.13  &1.11  &-0.020  &1.05  \\
$\Xi^-_{5c}$,$\Xi'^-_{5c}$,$\Xi^0_{5b}$,$\Xi'^0_{5b}$              &-0.14  &-0.22 &-0.054  &-0.30 \\
$\Xi^{--}_{5c}$,$\Xi^-_{5b}$                                       &-0.15  &-1.54 &-0.089  &-1.66 \\
 \hline
\end{tabular}
\end{center}
\caption{Numerical results of heavy pentaquark magnetic moments in
unit of $\mu_N$ in Karliner and Lipkin's model.
}\label{heavy-lipkin-mag}
\end{table}

It is interesting to compare the phenomenology of the above two
clustered quark models. There is only a $SU(3)$ flavor anti-sextet
for heavy pentaquarks in Jaffe and Wilczek's model. In contrast,
an additional flavor triplet exists in Karliner and Lipkin's
diquark and triquark model. The masses and magnetic moments of
these triplet pentaquarks are equal to their anti-sextet partners
in KL's model. For the light pentaquarks, all the above two
clustered quark models predicted the existence of an anti-decuplet
and an accompanying octet.

Another dramatic difference lies in the prediction of the magnetic
moments of the $J^P=\frac{1}{2}^+$ heavy sextet pentaquarks.  The
magnetic moments of the $J^P=\frac{1}{2}^+$ heavy sextet is tiny
and much smaller in KL's model than those in JW's model. There is
strong cancellation between the magnetic moment of the triquark
and the orbital magnetic moment in a $J^P=\frac{1}{2}^+$ sextet
when they are combined by Clebsch-Gordan coefficients. This
cancellation is almost exact for bottom pentaquarks. For the
magnetic moments of the $J^P=\frac{3}{2}^+$ sextet, there is also
significant difference between KL's model and JW's models.

The third difference is that the masses of the pentaquarks in KL's
model are about 300 MeV larger than those in JW's model. The
reason is that Karliner and Lipkin assumed that the diquark
(triquark) mass is simply the sum of its constituents. In
contrast, the diquark mass is made much lower than the sum of the
two quarks through the strong correlation between light quarks
when they are in anti-symmetric configuration in JW's model
\cite{jaffe}.

The heavy pentaquark masses and their strong, semi-leptonic and
non-leptonic decays were also discussed in
\cite{kingman,he,cheng,chiu2,huntheavy}. Besides the positive
parity states, it was pointed out in Ref. \cite{wise} that the
lowest heavy pentaquark carries negative parity in the diquark
model. Interested readers may consult these references for
details.

\section{Chiral Lagrangian formalism for pentaquarks}

The approximate chiral symmetry and its spontaneous breaking have
played an important role in hadron physics. Through the nonlinear
realization of spontaneous chiral symmetry breaking, we may study
the interaction between the chiral field and hadrons. The nonzero
current quark mass breaks the chiral symmetry explicitly.
Generally speaking, chiral symmetry provides a natural framework
to organize the hadronic strong interaction associated with the
light quarks. Chiral Lagrangian formalism has been used to discuss
pentaquarks in Refs. \cite{mehen,lee,zhanga,ko,liu-cpt}.
Interested readers are referred to these papers for the discussion
of pentaquark strong decay modes. In the following I only discuss
the mass splitting and selection rules arising from octet
pentaquark decays.

\subsection{Mass splitting}

The nonzero current quark mass induces mass splitting in the
multiplet. For example, these symmetry breaking terms for the
decuplet $P$ and negative parity octet $O_2$ are
\begin{eqnarray}\label{pp}
L_P &=& \alpha_P \,\overline{P} ( \xi m \xi + \xi^{\dagger} m
\xi^{\dagger} ) P,\\  \nonumber L_{O_2}&=& \alpha_{O_2}
\,\textrm{Tr}\big(d_2\overline{O_2}\{\xi m \xi + \xi^{\dagger} m
\xi^{\dagger},O_2\} + f_2 \overline{O_2} [\xi m \xi +
\xi^{\dagger} m \xi^{\dagger},O_2]\big) \nonumber\\  \nonumber &&
+\beta_{O_2} \, \textrm{Tr}(\overline{O_2}O_2) \textrm{Tr} ( m
\Sigma + \Sigma^{\dagger} m ).
\end{eqnarray}

Expanding Eq. (\ref{pp}), we get the mass splittings $\Delta
m_i\equiv m_i-m_{penta}$ for pentaquark anti-decuplet $P$
\begin{subequations}
\begin{eqnarray} \nonumber
\Delta m_\Theta & = & 2 \alpha_P m_s,
\\ \nonumber
\Delta m_{N_{10}} & = & \frac23\alpha_P \left( \hat{m} + 2 m_s
\right),
\\ \nonumber
\Delta m_{\Sigma_{10}} & = & \frac23 \alpha_P \left( 2 \hat{m} +
m_s \right),
\\ \nonumber
\Delta m_{\Xi_{10}} & = &  2~ \alpha_P \hat{m}.
\end{eqnarray}
\end{subequations}
Or
\begin{subequations}
\begin{eqnarray} \nonumber
m_{N_{10}}-m_{\Sigma_{10}}&=&m_\Theta-m_{N_{10}},\\ \nonumber
m_{\Sigma_{10}}-m_{\Xi_{10}}&=&m_{N_{10}}-m_{\Sigma_{10}}.
\end{eqnarray}
\end{subequations}

Similarly, for the $J^P={1\over 2}^-$ pentaquark octet $O_{2}$
\cite{zhanga}
\begin{subequations}
\begin{eqnarray} \nonumber
\Delta m_{N_{8_2}} & = & [2\beta_{O_2} + \alpha_{O_2} (d+f)]
(2\hat{m}) + [\beta_{O_2}+\alpha_{O_2}(d-f)] (2m_s),
\\ \nonumber
\Delta m_{\Sigma_{8_2}}&=&(\beta_{O_2}+\alpha_{O_2}d) (4\hat{m}) +
2 \beta_{O_2}m_s,
 \\ \nonumber
\Delta m_{\Xi_{8_2}}&=& [2\beta_{O_2} + \alpha_{O_2} (d-f)]
(2\hat{m}) + [\beta_{O_2}+\alpha_{O_2}(d+f)] (2m_s),
\\ \nonumber
\Delta m_{\Lambda_{8_2}}&=&(\beta_{O_2}+\frac13 \alpha_{O_2} d )
(4 \hat{m})+(\beta_{O_2}+\frac43\alpha_{O_2}d)(2m_s).
\end{eqnarray}
\end{subequations}
Hence we have the mass relation
\begin{equation}
2M_{N_8}+2M_{\Xi_8}=3M_{\Lambda_8}+M_{\Sigma_8}.
\end{equation}

\subsection{Selection rules from octet pentaquark decays}

In the framework of the diquark model, keeping explicit track of
the flavor indices of the two diquarks within the octet pentaquark
minimize the independent coupling constants and lead to some
selection rules caused by the "fall-apart" decay mechanism
\cite{maltman,close,lee,zhanga}. It's well-known that there are
two independent ways to couple three octets into a singlet, i.e.,
the D/F scheme. However, within the diquark model, when the octet
pentaquarks decay, there is only one independent coupling
constant. The underlying physics is quite simple. When $\Theta^+$
pentaquark falls apart and decays into a nucleon and kaon, the
$\bar s$ has to combine with a light quark to form a kaon.

In the following we use the $J^P={1\over 2}^-$ octet pentaquark
decay to illustrate this point. We denote a quark and anti-quark
by $q^i, {\bar q}_j$ where $i, j$ are the $SU(3)_F$ flavor
indices. Note that the flavor wave function of the $J^P={1\over
2}^-$ octet and singlet pentaquark arise from
\begin{equation}
A_{[ij]}\otimes\bar{q}_k=S\oplus O_{[ij,k]},
\end{equation}
where the indices $ij$ are antisymmetric, $A_{[ij]}$ is the
$\bf{3}_F$ diquark pair. $S$ is the pentaquark singlet whose
indices are contracted completely. $O_{[ij,k]}$ is the octet
representation. The index $k$ represents the antiquark which
contracts with one of the meson index.

For the interaction of the $J^P={1\over 2}^-$ pentaquark octet
$P$, nucleon octet $B$ and pseudoscalar meson octet $M$, we have
\cite{zhanga}
\begin{equation}
{\cal{L}}_8=g_8\epsilon_{ilm}\bar{O}^{[ij,k]}B^l_jM^m_k+H.c.,
\end{equation}
where $O_{[ij,k]}=\epsilon_{ljk}P^l_i-\epsilon_{lik}P^l_j$. We
present the Clebsch-Gordan coefficient of each interaction term in
Table \ref{selection-rule}.

\begin{table}[h]
\begin{center}
\begin{tabular}{cc|cc|cc|cc}\hline
$\Xi^-_8$             &  &   $\Xi^0_8$&       &$p_8$ && $n_8$&\\
\hline

$\Xi^-\pi^0$ & $\frac{1}{\sqrt{2}}$ & $\Xi^-\pi^+$ & 1 & $\Sigma^0 K^+$ &  $\frac{1}{\sqrt{2}}$  &$\Sigma^- K^+$ &  1\\
$ \Xi^0\pi^-$ & 1 & $\Xi^0\pi^0$ &-$\frac{1}{\sqrt{2}}$  & $\Sigma^+ K^0$ &  1 &$\Sigma^0 K^0$ &-$\frac{1}{\sqrt{2}}$\\
$\Xi^-\eta_0$ & $\frac{1}{\sqrt{6}}$ & $\Xi^0\eta_0$ &$\frac{1}{\sqrt{6}}$  & $p\eta_0$ &-$\frac{2}{\sqrt{6}}$ &$n\eta_0$ &-$\frac{2}{\sqrt{6}}$\\
$\Lambda K^-$ &-$\frac{2}{\sqrt{6}}$ & $\Lambda\bar{K^0}$ &-$\frac{2}{\sqrt{6}}$  &$\Lambda K^+$ & $\frac{1}{\sqrt{6}}$ &$\Lambda K^+$ & $\frac{1}{\sqrt{6}}$\\
 \hline
$\Sigma^0_8$             &  &   $\Sigma^+_8$&       &$\Sigma^-_8$ && $\Lambda_8$&\\
\hline
$\Sigma^+ \pi^-$&$\frac{1}{\sqrt{2}}$&$\Sigma^+\pi^0$&-$\frac{1}{\sqrt{2}}$&$\Sigma^-\pi^0$&$\frac{1}{\sqrt{2}}$ &$\Sigma^+\pi^-$&$\frac{1}{\sqrt{6}}$\\
$\Sigma^- \pi^+$&-$\frac{1}{\sqrt{2}}$&$\Sigma^0\pi^+$&$\frac{1}{\sqrt{2}}$&$\Sigma^0\pi^-$&-$\frac{1}{\sqrt{2}}$&$\Sigma^-\pi^+$&$\frac{1}{\sqrt{6}}$\\
$\Sigma^0 \eta_0$&$\frac{1}{\sqrt{6}}$&$\Sigma^+\eta_0$&$\frac{1}{\sqrt{6}}$&$\Sigma^-\eta_0$&$\frac{1}{\sqrt{6}}$ &$\Sigma^0\pi^0$&$\frac{1}{\sqrt{6}}$\\
$p K^-$&$\frac{1}{\sqrt{2}}$&$p \bar{K^0}$& 1 &$n K^-$& 1 &$p K^-$&$\frac{1}{\sqrt{6}}$\\
$n \bar{K^0}$&-$\frac{1}{\sqrt{2}}$&$\Lambda\pi^+$& $\frac{1}{\sqrt{6}}$ &$\Lambda\pi^-$& $\frac{1}{\sqrt{6}}$&$n \bar{K^0}$&$\frac{1}{\sqrt{6}}$\\
$\Lambda\pi^+$& $\frac{1}{\sqrt{6}}$ & & &&& $\Xi^- K^+$&-$\frac{2}{\sqrt{6}}$\\
 &   & & & & & $\Xi^0 K^0$&-$\frac{2}{\sqrt{6}}$\\
 &   & & & & & $\Lambda \eta_0$&-$\frac{1}{\sqrt{6}}$\\
 \hline
\end{tabular}
\end{center}
\caption{Coupling of the $J^P=\frac12^-$ pentaquark octet with the
usual baryon octet and the pseudoscalar meson octet. The universal
coupling constant $g_8$ is omitted.  } \label{selection-rule}
\end{table}

\section{Diquarks, pentaquarks and dibaryons}

The diquark is very similar to an anti-quark in many aspects
except charge and baryon number. This feature leads to deep
connection between pentaquarks and dibaryons which are composed of
three diquarks \cite{dib}.

For the $P=-$ dibaryons composed of three scalar diquarks with
$L=1$, its color wave function is anti-symmetric. Its spin wave
function is symmetric since diquarks are scalars. Bose statistics
requires the total wave function is symmetric. Hence the product
of the flavor and orbital wave function is anti-symmetric. Suppose
there is one orbital excitation between two diquarks: A and B. The
flavor wave function of the diquark pair A and B must be
symmetric, which is the same as in the $P=+$ pentaquarks. When the
orbital wave function is mixed symmetric (or anti-symmteric), the
flavor wave function must be mixed anti-symmteric (or symmetric).
This situation is very similar to the $L=1$ baryon multiplet in
the $SU(6)_{FS}$ $70_{FS}$ representation. The only difference is
that the diquark is a scalar. Simple group theory tells us that
the resulting $P=-$ dibaryons are in the $8_F$ representation.
Using the $\Theta^+$ mass as input, the negative baryons are
estimated to be slightly above two baryon threshold \cite{dib}.

Let's move on to those dibaryons which are composed of three
diquarks and have no orbital excitation. Three ${\bar 3}_c$
diquarks combine into a color singlet so their color wave function
is antisymmetric. Diquarks are scalars. They obey Bose statistics.
Their total wave function should be symmetric. Since there is no
orbital excitation between scalar diquarks, their spin and spatial
wave functions are symmetric. Hence their flavor wave function
must be totally anti-symmetric. I.e., the resulting dibaryon is a
$SU(3)_F$ singlet with positive parity, which is nothing but the H
dibaryon proposed by Jaffe long time ago \cite{jaffeold}. It's
interesting to note that the $J^P={1\over 2}^-$ pentaquark singlet
is very similar to the H dibaryon. From the available experimental
constraint on the H dibaryon binding energy, one may estimate the
singlet mass to be around $(1402\sim 1542)$ MeV, which is very
probably low-lying in the framework of diquark model \cite{dib}.

\section{Summary}\label{sec8}

The surprising discovery of pentaquarks shall be a milestone in
the hadron spectroscopy, if these states are further established
experimentally. Perhaps, a new landscape is emerging on the
horizon, of which we have only had a first glimpse through the
above experiments. Theorists have to answer the following
fundamental questions:
\begin{itemize}
\item  Is the world composed of quarks and gluons as we are
familiar with or a bunch of chiral solitons as indicated by the
chiral soliton model?

\item  Why are pentaquarks so different from conventional hadrons?
Does $\Theta^+$ carry positive or negative parity? Why is its
width extremely small? How to generate the low mass of $\Theta^+$
pentaquark?

\item  Does there exist strong correlation between light quarks?
Does it play an essential role in lowering the mass of the
pentaquark system?

\item  If so, what's the underlying dynamics leading to the strong
correlation? Is such a dynamics and correlation particular to the
pentaquark system or universal?

\item  Are there other "genuine" hadrons with valence quark
(anti-quark) number $N=4, 6, 7, 8, \cdots$, in which quarks do not
form two or more color-singlet clusters such as hadronic molecules
or nuclei?

\item  Is there an upper limit for $N$? In our universe, there may
exist quark stars where the quark number is huge. Is there a gap
in $N$ from pentaquark states to quark stars?
\end{itemize}
All these are very interesting issues awaiting further
experimental exploration, which will eventually deepen our
knowledge of the non-perturbative aspects of Quantum
Chromo-dynamics.

The author thanks Prof Yajun Mao for providing the first figure.
The author also thanks Prof W.-Y. P. Hwang and COSPA center at
National Taiwan University for the warm hospitality where part of
this project was done. This short review may be biased because of
the author's personal research interest. The literature of
pentaquarks has increased dramatically within one year. It is
impossible to cover all the important topics in this rapidly
developing field within this short review. The author apologize to
those whose papers and contributions were not mentioned in this
work. This project was supported by the National Natural Science
Foundation of China under Grant 10375003, Ministry of Education of
China, FANEDD and SRF for ROCS, SEM.


\end{document}